\documentclass[aps, a4paper,superscriptaddress,12pt,preprintnumbers,floatfix,nofootinbib,amsmath,amssymb]{revtex4}
\usepackage{amstext,amssymb}
\usepackage{amsmath}
\usepackage{graphicx}
\usepackage[hyperfootnotes=false]{hyperref}
\usepackage{xspace}
\usepackage{color}
\usepackage{units}
\usepackage{subfigure}
\usepackage{slashed} 

\newcommand{\bea}{\begin{eqnarray}}
\newcommand{\eea}{\end{eqnarray}}
\newcommand{\nn}{\nonumber}

\global\long\def\d{\partial}

\def\s1{\hat s}

\def\U1mt{U(1)_{L_\mu-L_\tau}}

\def\A{\mathcal{A}}

\def\SO10{\text{SO}(10)}

\def\SU{\,\text{SU}}

\def \epsilon {\varepsilon} 

\usepackage{longtable}
\usepackage{needspace}
\def\SO10{\text{SO}(10)}
\newcommand{\blue}[1]{{\color{blue} #1 }}

\begin{document}
\title{\hspace*{-1.0cm} Flavour anomalies and dark matter assisted unification in $SO(10)$ GUT}
\author{Purushottam \surname{Sahu}}
\email{purushottams@iitbhilai.ac.in}
\affiliation{Department of Physics, Indian Institute of Technology Bhilai, India}
\author{Aishwarya \surname{Bhatta}}
\email{aish.bhatta@gmail.com}
\affiliation{School of Physics, University of Hyderabad, Hyderabad 500046, India}
\author{Rukmani \surname{Mohanta}}
\email{rmsp@uohyd.ac.in}
\affiliation{School of Physics, University of Hyderabad, Hyderabad 500046, India}
\author{Shivaramakrishna \surname{Singirala}}
\email{krishnas542@gmail.com}
\affiliation{School of Physics, University of Hyderabad, Hyderabad 500046, India}
\author{Sudhanwa \surname{Patra}}
\email{sudhanwa@iitbhilai.ac.in}
\affiliation{Department of Physics, Indian Institute of Technology Bhilai, India}

\begin{abstract}
\vspace*{1cm}
With the recent experimental hint of new physics from flavor physics anomalies, combined with the evidence from neutrino masses and dark matter, we consider a minimal extension of SM with a scalar leptoquark and a fermion triplet. The scalar leptoquark with couplings to leptons and quarks can explain lepton flavor non-universality observables $R_K$, $R_{K^{(*)}}$, $R_{D^{(*)}}$ and $R_{J/\psi}$. Neutral component of fermion triplet provides current abundance of dark matter in the Universe. The interesting feature of the proposal is that the minimal addition of these phenomenologically rich particles (scalar leptoquark and fermion triplet) assist in realizing the unification of the gauge couplings associated with the strong and electroweak forces of standard model when embedded in the non-supersymmetric $SO(10)$ grand unified theory. We discuss on unification mass scale and the corresponding proton decay constraints while taking into account the GUT threshold corrections.
  
\end{abstract}
\maketitle
\newpage
\section{Introduction}
Standard model of particle physics (SM) beautifully explains the gauge theory of strong, electromagnetic and weak interactions, with all its predictions testified at current experiments including LHC. Still it is known that many observed phenomena like neutrino masses and mixing \cite{Bilenky:1999ge,Mohapatra:1979ia,Schechter:1980gr,Babu:1992ia,Super-Kamiokande:2005wtt,SNO:2002tuh,Super-Kamiokande:2016yck,T2K:2019efw,DayaBay:2012fng,DoubleChooz:2011ymz}, dark matter \cite{Zwicky:1937zza,PhysRev.43.147,Bertone:2004pz,Mambrini:2015vna}, matter anti-matter asymmetry \cite{Sakharov:1967dj,Kolb:1979qa,Fukugita:1986hr,Fritzsch:1974nn} and the recent flavor anomalies, see for example \cite{Bifani:2018zmi} and references therein, cannot be addressed within its framework. This motivates to explore other possible beyond standard model (BSM) frameworks which have the potential to address these unsolved issues of the SM. It is believed that the ultimate theory of  elementary particles might be an effective low energy approximation of some grand unified theory (GUT) or part of another theory at high scale.

Though most of the flavor observables go along with the SM, there are a collection of recent measurements in  semileptonic $B$ meson decays, involving $b \to s \ell \ell$ ($\ell=e,\mu)$ and $b \to c l \bar \nu_ l$ $(l=\mu,\tau)$ quark level transitions, that are incongruous with the SM predictions. The most conspicuous measurements, hinting the physics beyond SM are the lepton flavor universality violating parametes:  $R_{K}$ with a discrepancy of $3.1\sigma$ \cite{Aaij:2014ora, Aaij:2019wad, LHCb:2021trn, Bobeth:2007dw, Bordone:2016gaq}, $R_{K^{(*)}}$ with a disagreement at the level of $(2.1-2.5)\sigma$ \cite{Aaij:2017vbb, Capdevila:2017bsm}, $R_{D^{(*)}}$ with $3.08\sigma$ discrepancy \cite{HFLAV:2019otj, Na:2015kha, Fajfer:2012vx, Fajfer:2012jt} and $R_{J/\psi}$ with a deviation of nearly $2\sigma$ \cite{Aaij:2017tyk, Wen-Fei:2013uea, Ivanov:2005fd} from their SM predictions. Though the Belle Collaboration \cite{Abdesselam:2019lab, Abdesselam:2019wac} has also announced their measurements on $R_{K^{(*)}}$ in various $q^2$ bins, however these measurements have large uncertainties.  Besides the $R_{K^{(*)}}$ parameters,  the $P_5^\prime$ optimized observable disagrees with the SM at the level of $4\sigma$ in the $(4.3-8.68)~\rm {GeV}^2$ $q^2$-bin \cite{Aaij:2013qta, LHCb:2015svh, Belle:2016xuo} and the decay rate  of  $B \to K^* \mu \mu$ shows $3\sigma$ discrepancy \cite{Aaij:2014pli}. The branching ratio of $B_s \to \phi \mu \mu$ channel also disagrees with the theory at the level of $3\sigma$ \cite{Aaij:2013aln} in low $q^2$.

One of the possible explanation for these flavor  anomalies is the existence of leptoquarks (LQ) leading to the transitions  $b \to s \ell \ell$ and $b \to c l \bar \nu_l$. It is believed that LQs may lead to interesting new physics searches and could be the next big discovery at LHC. Since, by definition, LQ connecting both leptons and quarks simultaneously may have its origin from quark-lepton symmetry, Pati-Salam symmetry, $SO(10)$ and other grand unified models (GUT). in the present work, we wish to study LQ assisted gauge coupling unification of the fundamental forces described by the SM. In the present work, the idea is to construct a TeV scale extension of SM in order to explain the experimental hints of new physics in recently observed flavor  anomalies within the framework of non-supersymmetric $SO(10)$ grand unified theory while simultaneous addressing neutrino mass and dark matter. The important feature of the model is that inclusion of a scalar LQ and a fermion triplet DM at few TeV scale on top of SM leads to successful unification of SM gauge couplings.

In the context of GUT, the popular models are $SU(5)$ \cite{Georgi:1974sy}, $SO(10)$ \cite{Pati:1974yy,Fritzsch:1974nn,PhysRevD.48.264,Senjanovic:1975rk,Clark:1982ai,Altarelli:2013aqa,Dueck:2013gca,Meloni:2014rga,Meloni:2016rnt,Preda:2022izo,Chakrabortty:2017mgi,Bandyopadhyay:2017uwc} and $E_6$\cite{Gursey:1975ki,Shafi:1978gg,Nandi:1985uh,Stech:2003sb,Huang:2014zba,Dash:2020jlc,Dash:2019bdh,Dash:2021suj,Bandyopadhyay:2019rja}, where many of the unsolved issues of the SM can be addressed. In most of the literature, it is found that all GUTs without any intermediate symmetry breaking and in the absence of supersymmetry, fails to unify the gauge couplings corresponding to three fundamental forces as described by SM. Few attempts were successful in gauge coupling unification by adding extra particles on top of SM spectrum at a higher scale. With this idea, we explore a simplified extension of SM at few TeV scale, which can be successfully embedded in a non-supersymmetric $SO(10)$ GUT. The key feature of the work is that the extra particles, isospin triplet fermion and scalar leptoquark (SLQ) which are originally motivated to unify the gauge couplings, can simultaneously address the dark matter of the Universe and flavor anomalies. While examining the gauge coupling unification it is observed that the unification scale and inverse fine structure constant are in conflict with proton decay prediction. In order to satisfy the proton decay limits, we propose the presence of super heavy particles including scalars, fermions and gauge bosons sitting at GUT scale, which can modify the unification scale and the inverse fine structure constant can be explained through one-loop GUT threshold effects\cite{Mohapatra:1992jw,Hall:1980kf,Babu:2015bna,Parida:2016hln,Schwichtenberg:2018cka,Chakrabortty:2019fov,Dash:2020jlc}.


The structure of the paper is as follows. In section-II, a realistic TeV scale extension of SM with scalar LQ, fermionic triplet DM and its embedding in non-supersymmetric $SO(10)$ GUT is proposed. Section-III discusses the implications of GUT threshold corrections to gauge coupling constants and unification mass scale in order to comply with the current bound on proton decay. In section-IV, we comment on fermion masses and mixing including the light neutrino masses via type-I seesaw. Addressing of flavor anomalies with scalar LQ is presented in section-V. Section-VI discusses the role of fermion triplet as DM candidate, which was originally motivated for gauge coupling unification. We conclude our results in section-VII.

%
\section{Leptoquark and DM assisted gauge coupling unification}
It has been established in a number of investigations \cite{FUKUYAMA_2013,Frigerio:2011zg,Alonso:2014zka, Dorsner:2011ai,Chang:1984qr,Bertolini_2009} that non-supersymmetric grand unified theories including $SO(10)$ GUT can provide successful gauge coupling unification with either an intermediate symmetry or inclusion of extra particles. At the same time, the inability of SM to explain the non-zero neutrino masses, dark matter and recent flavour anomalies requires to explore possible SM extensions. Combining these two ideas, we wish to consider a minimal extension of SM and examine how the unification of gauge couplings are achieved with the minimal extension of SM with a scalar leptoquark $R_2 (3_C,2_L,7/6_Y)$ and a fermion triplet $\Sigma (1_C,3_L,0_Y)$ around TeV scale by embeding the set up in a non-supersymmetric $SO(10)$ GUT with the following symmetry breaking chain,
\begin{eqnarray}
SO(10)
&&	\stackrel{M_U}{\longrightarrow} SU(3)_C\otimes SU(2)_L\otimes U(1)_{Y} \quad \left(\mbox{with}~ R_2, \Sigma \right) \nonumber\\ 
&&	\stackrel{M_I}{\longrightarrow}SU(3)_C\otimes SU(2)_L\otimes U(1)_{Y}   \nonumber \\
&&	\stackrel{M_Z}{\longrightarrow} SU(3)_C\otimes U(1)_{Q}\;.
\label{Chain}
\end{eqnarray}
Instead of introducing an intermediate symmetry between $SO(10)$ and SM, we take an intermediate mass scale ($M_I$) and two new fields $R_2$ and $\Sigma$ are included.

It is also important to note that scalar leptoquarks can arise naturally in grand unified theories like Pati-Salam (PS) model based on the gauge group $ SU(4)_C\otimes SU(2)_L\otimes SU(2)_{R}$ \cite{Pati:1973uk,Pati:1974yy}. PS model which was originally motivated for quark lepton mass unification already accommodates all scalar LQs mediating interesting B-meson anomalies while keeping the relevant LQ mass to few TeV scale. The issue with simple SM extension with LQs is that it may lead to proton decay, which requires additional symmetry to stabilize the proton. However, the leptoquarks originated from PS symmetry mediate B-physics anomalies but do not cause proton decay. With this motivation we can also consider other novel symmetry breaking chain as
\begin{eqnarray}
SO(10)
&&	\stackrel{M_U}{\longrightarrow} G_{422 (D)} 
	\stackrel{M_I}{\longrightarrow} G_{321} \stackrel{M_Z}{\longrightarrow} G_{31} 
\end{eqnarray}
where, the used notations are,
\begin{eqnarray}
G_{422}
&&	=  SU(4)_C \otimes SU(2)_L\otimes SU(2)_{R} \nonumber\\
G_{321}
&&	= SU(3)_C\otimes SU(2)_L\otimes U(1)_{Y} \nonumber\\ 
G_{31} && = SU(3)_C\otimes U(1)_{Q}\;.
\end{eqnarray}

The first stage of symmetry breaking i.e, $SO(10) \rightarrow G_{422}$ is achieved by giving non-zero vev to $G_{422}$ singlets in $54_H$ (Case A) or $210_H$ (Case B) at unification scale $M_U$. It is to be noted that the vev assignment to singlet $\langle(1,1,1)\rangle$ belonging to $54_H$ is even under D-parity. Therefore D-parity is not broken while the vev assignment of the singlet $\langle(1,1,1)\rangle$ belonging to $210_H$ is odd under D-parity. In the next stage symmetry breaking PS to SM gauge group i.e, $ G_{422} \rightarrow G_{321} $ at $M_I$ energy scale is achieved by giving non-zero vev to SM singlet contained in   $\Delta_{R}(\overline{10}_{4c}, 1_{2L},3_{2R})$ of $126_H$.
The final stage of the symmetry breaking $ G_{321} \rightarrow G_{31}$ is achieved by the SM Higgs doublet  contained in $\phi(1_{4c},2_{2L},2_{2R})$ of $10_H$. Here $M_I$ is the energy scale at which the Pati-Salam symmetry is broken into the SM, which is the mass scale of these Pati-Salam
multiplets. All the remaining fields are assumed to be heavy at the unification scale $M_U$. In order to maintain a complete left-right symmetry for Case-A, we added   Pati-Salam multiplets $\Delta_{L}(\overline{10}_{4c}, 3_{2L},1_{2R})$ and $\Sigma_{R}(\overline{1}_{4c}, 1_{2L},3_{2R}),$  at $M_I$ .  In each energy scale the particle content and the corresponding beta coefficients are given in Table \ref{tab:PS}.

\begin{table}[!htb]
\caption{\label{tab:PS} The particles (scalars and fermions) content and the Beta coefficients in the breaking intervals for case A and B. Case A is valid for intermediate PS symmetry with discrete D-parity invariance with the presence of extra fields marked in blue.}
\centering
\begin{tabular}{lcc}
\hline \hline Interval & Particle content for case $A\left(B\right)$ & Beta coefficients $A\left(B\right)$ \\
\hline &Scalars & \\
$M_{I}-M_{U}$ & $\phi_1(1,2,2), \phi_2(1,2,2)$ & {$\left[b_{4c}, b_{2L}, b_{Y}\right]=\left[\frac{2}{3}\left(\frac{-7}{3}\right), \frac{31}{3}\left(\frac{11}{3}\right), \frac{31}{3} \left(\frac{27}{3}\right)\right]$} \\
& $R_2(15,2,2) , \Delta_{R}(\overline{10}, 1,3),$ & \\
& \blue {$\Delta_{L}(10, 3,1) $} (Case-A) & \\
& Fermions & \\
&$\Psi_L(4,2,1), \Psi_R(\overline{4},1,2)$ & \\
& $\Sigma_L(1,3,1), \blue{\Sigma_R(1,1,3)}$ & \\
& Scalars & \\
$M_Z-M_{I}$ & $H(1,2,1/2) , R_2(3,2,7/6)$  & {$\left[b_{3c}, b_{2L}, b_{Y}\right]=\left[\frac{-20}{3}, \frac{-4}{3}, \frac{86}{15}\right]$} \\
& Fermions & \\
& $ Q_L(3,2,1/6) , u_R (3,1,2/3),d_R(3,1,-1/3)$ & \\
& $L_L(1,2,-1/2),e_R (1,1,-1) $ & \\
&$N_R(1,1,0) ,\Sigma(1,3,0)$ & \\
\hline \hline
\end{tabular}
\end{table}


In our analysis, the required non-trivial degrees of freedom with fermion
triplet dark matter and a scalar leptoquark at TeV scale can lead to gauge coupling unification.
The inclusion of Pati-Salam intermediate symmetry only safeguards from rapid proton decay
due to scalar leptoquarks at TeV scale, but does not lead to any significant modification to the
unification mass scale and gauge coupling unification.

The known SM fermions plus additional sterile neutrinos are contained in $16_F$ spinorial representation of $SO(10)$ as follows
\begin{eqnarray}
16_F&=&Q_L\,(3,2,1/6) + u_R\,(3,1,2/3) + d_R\,(3,1,-1/3) 
\nonumber \\
&&+ L_L \,(1,2,-1/2) + e_R\,(1,1,-1) 
+N_R(1,1,0)\nonumber \\
&=& 15_F (\mbox{SM Fermions}) + N_R\, \mbox{(sterile neutrino).}
\end{eqnarray}
Thus it is obvious that the $16_F$ spinorial representation provides unification in the matter sector. The presence of sterile neutrinos in $16_F$ provides sub-eV scale of neutrino masses via type-I seesaw \cite{Minkowski:2015pya,Ohlsson:2019sja,Akhmedov:2003dg} and also explains matter anti-matter asymmetry via leptogenesis\cite{Fong:2014gea,DiBari:2021fhs,Bodeker:2020ghk,Xing:2020ald,Buchmuller:1996pa,Nezri:2000pb,Buccella:2001tq,Branco:2002kt,DiBari:2008mp,DiBari:2010ux,Buccella:2012kc,DiBari:2014eya,DiBari:2015oca,DiBari:2017uka,DiBari:2020plh,Vives:2005ra,DiBari:2005st,Abada:2006fw,Abada:2006ea,Mummidi:2021anm}. The $10_H$ representation of $SO(10)$ GUT contains SM Higgs field $\phi(1_C,2_L, 1/2_Y)$ which is essential for electroweak symmetry breaking. The SM gauge bosons including $8$ gluons ($G^a_\mu$), three weak gauge bosons $W^+_\mu, W^-_\mu, Z_\mu$ and photon are contained in adjoint representation $45_V$ of $SO(10)$. 

The evolution of gauge coupling constants $g_i (\mu)$ ($i = 3_C, 2_L, Y$) using standard renormalization group equations (RGEs)~\cite{Georgi:1974yf} is given by,
\begin{equation}
\mu\,\frac{\partial g_{i}}{\partial \mu}=\frac{b_i}{16 \pi^2} g^{3}_{i}+
\frac{1}{(16 \pi^2)^2}\, \sum_{j} B_{ij} g^3_{i} g^2_{j}\, .
\label{rge-coupl}
\end{equation}
The solutions can be derived in terms of inverse coupling constant, valid from $\mu$ to the intermediate scale $M_I$ (with $M_I > \mu$) as,
\begin{eqnarray}
\frac{1}{\alpha_{i}(\mu)} &=& \frac{1}{\alpha_{i}(M_I)} + \frac{b_i}{2 \pi}\,  \text{\large ln}\left(\frac{M_I}{\mu} \right)+\frac{1}{8 \pi^2}\, \sum_{j} B_{ij} \int^{M_I}_{\mu} \alpha_j(\mu) \frac{d\mu}{\mu} \, .
\label{rge-alphainv}
\end{eqnarray}
Here, $\alpha_{i}=g^2_{i}/(4 \pi)$ and $b_i$ ($B_{ij}$) is the one (two)-loop beta coefficients in the mass range $M_Z-M_I$ and $M_I-M_U$ which are presented in Table.\ref{tab:beta}. $M_Z$ stands for electroweak scale, $M_I$ is intermediate scale and $M_U$ represents unification scale. \\
\begin{longtable}{ll c c}
\caption{\label{tab:beta} Beta coefficients at one-loop and two-loop levels.}
\endhead
\hline\hline
Mass Range &   1-loop level  & 2-loop level \\[1mm]
\hline
$M_Z-M_I$   &  $b_i = (-7, -\frac{19}{6}, \frac{41}{10})$ & $B_{ij} = \begin{pmatrix}
-26 & \frac{9}{2} & \frac{11}{10} \\
12 & \frac{35}{6} & \frac{9}{10} \\
\frac{44}{5} & \frac{27}{10} & \frac{199}{50}
\end{pmatrix}$\\
$M_I-M_U$   &  $b_i^\prime = (\frac{-20}{3}, \frac{-4}{3}, \frac{86}{15})$ & 
$B_{ij}^\prime = \begin{pmatrix}
    \frac{-56}{3}           &  \frac{15}{2}            &  \frac{131}{30}          \\
        20                &   \frac{86}{3}              &  \frac{29}{5}          \\
  \frac{524}{15} &   \frac{87}{5}           &  \frac{3721}{150}                  
\end{pmatrix}$ \\  
\hline
\end{longtable}

\begin{figure}[htb!]
\centering
\includegraphics[scale=0.65]{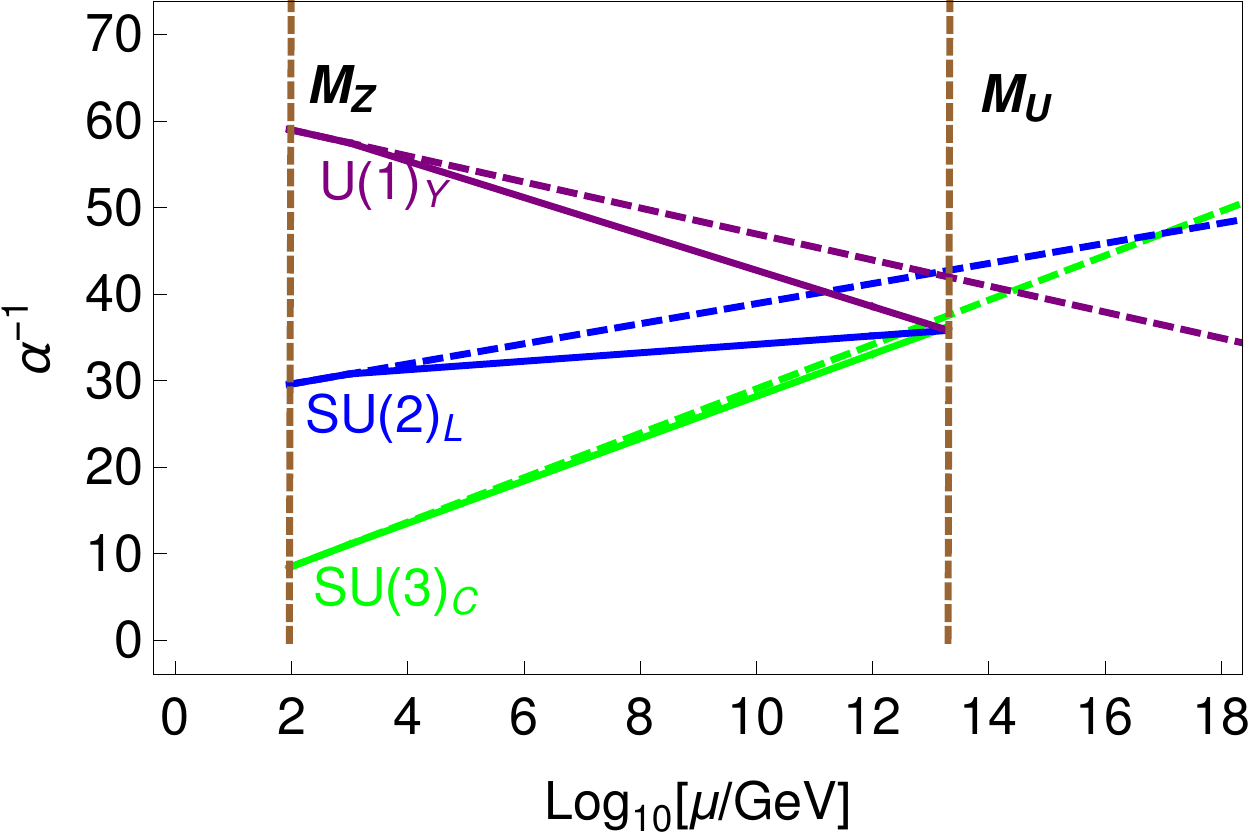}
\caption{Evolution of the gauge coupling constants of SM gauge symmetry, where dashed lines are contributions from SM particle content and solid lines correspond to RGEs with SM plus $R_2$ and $\Sigma$. The vertical dotted lines from left-right are representing symmetry breaking scales $M_Z$ as electroweak scale and $M_U$ as unification scale.}
\label{plot:unifn_wo_threshold}
\end{figure}

We skip the discussion RG evolution of gauge coupling constants with two loop effects. While the one-loop RGEs from mass scale $M_Z$ to $M_I$ and $M_I$ to $M_{U}$ are read as follows,
\begin{eqnarray}
\alpha^{-1}_{i} (M_Z)&&=\alpha^{-1}_{i} (M_I) + \frac{b_{i}}{2 \pi} {\large \ln}\left(\frac{M_I}{M_Z}\right), \,\nonumber \\
\alpha^{-1}_{i} (M_I)&&=\alpha^{-1}_{i} (M_U) + \frac{b^\prime_{i}}{2 \pi} {\large \ln}\left(\frac{M_U}{M_I}\right).
\label{eqn:alphaMI}
\end{eqnarray}
Simplifying RGEs, we obtain the analytic solution for unification mass scale as 
\begin{eqnarray}
&&{\large \ln}\left(\frac{M_U}{M_Z}\right)= 
 \frac{A_{I} D_{W}-B_{I} D_{S}}{B_{U} A_{I}-B_{I} A_{U}}, 
 \label{MU-1loop-wo-th} 
\end{eqnarray}
where, the parameters $D_{S}$ and $D_{W}$ are given by
\begin{equation}
D_S= 16\pi \left[\alpha_S^{-1}(M_Z)-\frac{3}{8}\alpha^{-1}_{\rm em}(M_Z) \right]\,,~~D_W= 16\pi \left[\sin^2\theta_W-\frac{3}{8} \right] \alpha^{-1}_{\rm em}(M_Z).
\end{equation}
While all other parameters are expressed in terms of one-loop beta coefficients as
\begin{eqnarray}
&&A_I= \Bigg[\Big(8 b_{3C}-3 b_{2L}-5 b_{Y} \Big)
-\Big(8 b_{3C}^{'}-3 b_{2L}^{'}-5b_{Y}^{'} \Big) \Bigg],~~ A_U= \left (8 b_{3C}^{'}-3 b_{2L}^{'}-5b_{Y}^{'} \right), \nonumber \\
&&B_I=\left[\left (5b_{2L}-5b_{Y} \right)-\left (5b_{2L}^{'}-5b_{Y}^{'} \right) \right], ~~B_U= \Big(5\,  b^\prime_{2L}  - 5\,  b^\prime_{Y}  \Big).
\end{eqnarray}
Using the experimental values of $\alpha_{\rm em}$, $\alpha_{\rm S}$, $M_{Z}$ and Weinberg mixing angle \cite{ParticleDataGroup:2014cgo,Mou:2017sjf}, the estimated values of  unification mass scale and inverse GUT coupling constant are given by 
\begin{equation}
M_U= 10^{13.27}\,\,\mbox{GeV and} \quad \alpha^{-1}_U=36.287. 
\label{eq:unifscale and coup}
\end{equation}

The evolution of gauge couplings for $SU(3)_C$, $SU(2)_L$ and $U(1)_Y$ gauge groups are displayed in Fig.\ref{plot:unifn_wo_threshold}. Here, the dashed (solid) lines correspond to SM contribution (SM plus $R_2$ and $\Sigma$ contributions). The purple line refers to inverse fine structure constant for $U(1)_Y$ group while blue and green lines correspond to $SU(2)_L$ and $SU(3)_C$ groups respectively. It is evident that the SM predictions with dashed lines demonstrate that there is no such gauge coupling unification. However, with the inclusion of extra particles on top of SM at TeV scale, evolution of gauge couplings begin to deviate from the SM results and provide successful gauge coupling unification of weak, electromagnetic and strong forces. 
\begin{figure}[h]
\centering
\includegraphics[scale=1.5]{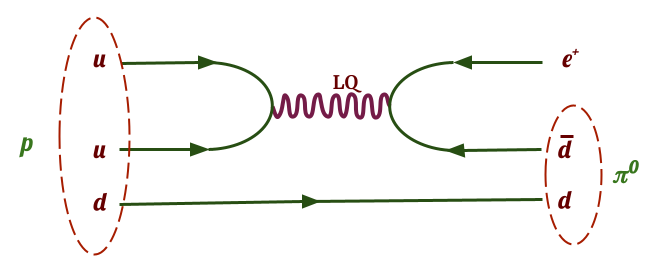}
\caption{Proton decay mediated by leptoquark gauge bosons contained in adjoint representation of $SO(10)$.}
\label{plot:pdecay}
\end{figure}
\subsection{Prediction of proton lifetime}
The interesting feature of grand unified theories is that they can have a robust prediction on proton decay with the presence of exotic interactions mediated by super heavy gauge bosons and scalars. Most commonly discussed gauge boson mediated proton decay arises from the covariant derivative of the fermions in $16_F$ with the gauge bosons contained in $45_V$ of $SO(10)$, leading to the interaction between quarks and leptons. In our framework, we assume that dominant contributions to proton decay to a neutral pion and a positron comes from the mediation of leptoquark gauge bosons in $45_V$.  For simplicity, we neglect the contributions form other super heavy particles. 

The gauge boson mediated proton life time in the process $p \to e^+ \pi^0$ ~\cite{Babu:1992ia, Bertolini:2013vta, Kolesova:2014mfa,Parida:2016hln,Meloni:2019jcf,Ibanez:1984ni, Buras:1977yy, BhupalDev:2010he,Chakrabortty:2019fov,Babu:2015bna} shown in Fig. \ref{plot:pdecay}, is given by
\begin{eqnarray}  
\tau_p={\Gamma}^{-1}\left( p\rightarrow \pi^0 e^+ \right) &=&
 \frac{64 \pi f^2_\pi}{m_p} \left(\frac{M^4_U}{g^4_U} \right)
\times \frac{1}{|{A_L}|^2 |\overline{\alpha}_H|^2 \left(1 + \mathcal{F} + \mathcal{D} \right)^2 {\mathcal{R}}} .
\label{eq:protonlife}
\end{eqnarray}
Here,  $g_U$ is the GUT scale coupling related with fine structure constant as $\alpha_U = g_U^2/4\pi$ and the predicted unification mass scale $M_U$ is the typical mass scale of all the super heavy particles. The other parameters, like $m_p$ stands for proton mass, $f_\pi$ is the pion decay constant ${\mathcal{R}}$ is the renormalization factor,  $ {\mathcal{R}}=\left({A_{SL}}^2+{A_{SR}}^2\right)
\left(1+ |{V_{ud}}|^2\right)^2\ $ with $V_{ud}$ being the CKM-matrix element. Defining ${A^2_R} =  {A^2_{L}} \left({A^2_{SL}} + {A^2_{SR}} \right)$ and $\alpha_H = \overline{\alpha}_H \left(1 + \mathcal{F} + \mathcal{D} \right)$, the proton lifetime is modified as follows
\begin{eqnarray}  
\tau_{p\rightarrow \pi^0 e^+} &=&
 \frac{4}{\pi}  \left(\frac{ f^2_\pi}{m_p}\right) \left(\frac{M^4_U}{\alpha^2_U} \right)
\frac{1}{\alpha^2_H {A^2_R} \left(1+ |{V_{ud}}|^2\right)^2}  \,.
\label{lifetime-proton-modified}
\end{eqnarray}
In the present model, the long distance enhancement factor is  $A_L\simeq 1.25$ while the short distance renormalization factors are $ A_{SL} = 2.46$ and $A_{SR}=2.34$. Using $\alpha_H = 0.012$ GeV$^3$ and the estimated values of $M_U$ and $\alpha_U$, the proton lifetime is found to be 
\begin{equation}
\tau_p = 4.16925 \times 10^{24} ~{\rm yrs}.
\end{equation}
This prediction is well below the current bound set by Super-Kamiokande \cite{Miura:2016krn} ($\tau^{\rm SK}_p > 1.6 \times 10^{34}\, \mbox{yrs}$) and Hyper-Kamiokande \cite{Abe:2011ts,Yokoyama:2017mnt} ($\tau_p^{\rm HK2025} > 9.0 \times 10^{34}\, \mbox{yrs}$). The gravitational corrections arising from higher dimensional operators or GUT threshold effects can enhance the unification mass scale $M_U$ and the proton lifetime, consistent with the experimental bounds. 

In the next section, we will estimate the one-loop GUT threshold contributions to evolution of gauge coupling constants starting from derived unification mass scale $M_U$. As a result, we get a threshold corrected unification mass scale $M^{ TH}_U$ and also examine whether the corrected proton lifetime is in agreement with the experimental constraints.
\section{GUT threshold predictions on unification scale and proton decay}
The idea of one-loop GUT threshold corrections is to shift the values of SM gauge couplings at $M_U$ with the  presence of super heavy particles. For illustration, let us consider the minimal $SO(10)$ Higgs representation as $10_S \equiv \phi (1,2,1/2) + S_1(1,2,-1/2) + S_2(3,1,-1/3) + S_3(3,1,1/3)$. Since, $\phi$ is utilized at low scale for electroweak symmetry breaking, all other scalars are considered as super heavy scalars and may contribute to the one-loop GUT threshold corrections. The same argument can be applied to other scalars/fermions/gauge bosons contained in different representation of $SO(10)$ presented in the Table. \ref{tab:so10-superheavy}.
\begin{table}[!htb]
\caption{ Super heavy scalars, fermions and vector bosons contributing to the GUT threshold corrections.}
\label{tab:so10-superheavy}
\centering
\begin{tabular}{l l l}
\hline\hline
&$SM$ & $G_{321}$  \\
\hline
Scalars & $10_S$ & $S_1(\pmb{1},\pmb{2}, -1/2)$, $S_2(\pmb{3},\pmb{1}, -1/3)$,  $S_3(\overline{\pmb{3}},\pmb{1}, 1/3)$ , 
\\
& $126_S$ &$2 S_4(\pmb{3},\pmb{1}, -1/3)$, $ S_5(\overline{\pmb{3}},\pmb{1}, 1/3)$, $S_6(\pmb{1},\pmb{3}, 1)$, $S_7(\overline{\pmb{6}},\pmb{3}, -1/3)$ , $S_8(\pmb{1},\pmb{1}, 0)$\\
& &$S_9(\pmb{1},\pmb{1}, -1)$, $S_{10}(\pmb{1},\pmb{1}, -2)$, $S_{11}(\pmb{3},\pmb{1}, 2/3)$, $S_{12}(\pmb{3},\pmb{1}, -4/3)$, $S_{13}(\pmb{6},\pmb{1}, 4/3)$\\
& &$S_{14}(\pmb{6},\pmb{1}, 1/3)$, $S_{15}(\pmb{6},\pmb{1}, 2/3)$, $S_{16}(\pmb{1},\pmb{2}, 1/2)$, $S_{17}(\pmb{1},\pmb{2}, -1/2)$, $S_{18}(\pmb{3},\pmb{2}, 1/6)$ \\
& & $S_{19}(\overline{\pmb{3}},\pmb{2}, -7/6)$, $S_{20}(\overline{\pmb{3}},\pmb{2}, -1/6)$, $S_{21}(\pmb{8},\pmb{2}, 1/2)$, $S_{22}(\pmb{8},\pmb{2}, -1/2)$, $S_{23}(\overline{\pmb{3}},\pmb{3}, 1/3)$ \\
\hline
Fermions & ${16}_F$ & \\
 & ${45}_F$ &$F_1(\pmb{1},\pmb{1}, 1)$, $2 F_2(\pmb{1},\pmb{1}, 0)$, $F_3(\pmb{1},\pmb{1}, -1)$, $F_4(\pmb{3},\pmb{2}, 1/6)$, $F_5(\pmb{3},\pmb{2}, -5/6)$\\
 & & $F_6(\overline{\pmb{3}},\pmb{2}, 5/6)$, $F_7(\overline{\pmb{3}},\pmb{2}, -1/6)$, $F_8(\pmb{3},\pmb{1}, 2/3)$, $F_9(\overline{\pmb{3}},\pmb{1}, -2/3)$, $F_{10}(\pmb{8},\pmb{1}, 0)$\\
\hline
Vectors & $45_V$ & $V_1(\pmb{1},\pmb{1}, 1)$, $ V_2(\pmb{1},\pmb{1}, 0)$, $V_3(\pmb{1},\pmb{1}, -1)$, $V_4(\pmb{3},\pmb{2}, 1/6)$, $V_5(\pmb{3},\pmb{2}, -5/6)$\\
 & & $V_6(\overline{\pmb{3}},\pmb{2}, 5/6)$, $V_7(\overline{\pmb{3}},\pmb{2}, -1/6)$, $V_8(\pmb{3},\pmb{1}, 2/3)$, $V_9(\overline{\pmb{3}},\pmb{1}, -2/3)$\\
\hline\hline
\end{tabular}
\end{table}
\subsection{Analytic formula for threshold corrections}
\begin{figure}[h]
\centering
\includegraphics[scale=0.65]{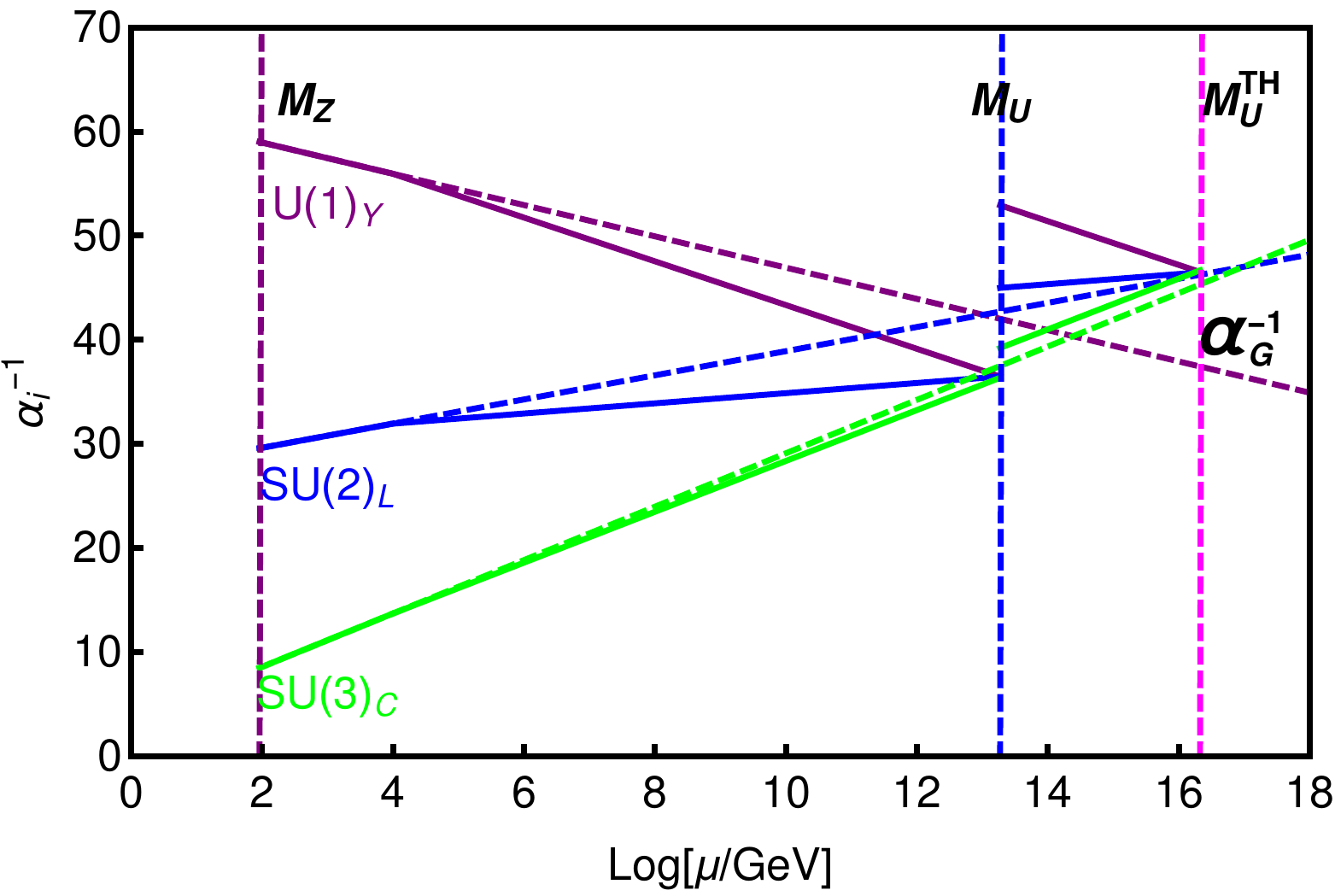}
\caption{Unification plot for all three gauge couplings, where the one-loop effects are displayed upto the mass scale $M_U$ while the threshold effects are shown in the mass range $M_U-M_U^{TH}$.}
\label{plot:unifn_threshold}
\end{figure}

The matching condition at a given symmetry breaking scale $\mu$, by including one-loop GUT threshold corrections is given by~\cite{Hall:1980kf,Parida:2016hln,Chakrabortty:2019fov,Babu:2015bna,Schwichtenberg:2018cka}
\begin{eqnarray}
 \alpha^{-1}_{D} (\mu)=\alpha^{-1}_{P} (\mu)-\frac{{\lambda_{D}(\mu)}}{12\pi},
\end{eqnarray}
where $\alpha_{P}^{-1}(\mu)$ and $\alpha_{D}^{-1}(\mu)$  denote the inverse coupling constant corresponding to the parent and daughter gauge groups. The parent gauge symmetry gets spontaneously broken down to the daughter gauge group at the mass scale $\mu=M_U$, where, the parent group is a simple $SO(10)$ and the daughter one is a product of different gauge symmetries i.e, $SU(3)_C \times SU(2)_L\times U(1)_Y$. In the present model, the matching conditions for all the inverse gauge couplings of SM at $M_U$ are read as
\begin{eqnarray}
 \alpha^{-1}_{i} (M_U)=\alpha^{-1}_{U} (M_U)-\frac{{\lambda_{i}(M_U)}}{12\pi}.
\end{eqnarray}
The threshold parameter ${\lambda_{i}}$ is a sum of individual contributions due to the presence of super heavy scalars, fermions and vector bosons (or gauge bosons) with masses $M_S$, $M_F$ and $M_V$ respectively at GUT scale, is given by
\begin{eqnarray}
&&\lambda_i(M_U) = \lambda^S_i(M_U) + \lambda^F_i(M_U)+\lambda^V_i(M_U),
\end{eqnarray}
where, 
\begin{eqnarray}
&& \lambda^S_i (M_U)= \sum_{j} 2\,k \mbox{\large Tr} \left[t_{i}^2(S_j) \large {\large \ln }\left(\frac{M_{S_j}}{M_U}\right)\right],\nonumber \\
&& \lambda^F_i(M_U)= \sum_{j} 8\,\kappa \mbox{\large Tr} \left[t^2_i(F_j) \large {\large \ln }\left(\frac{M_{F_j}}{M_U}\right)\right],\nonumber \\
&& \lambda^V_i (M_U)= \mbox{\large Tr} \left[t_i^2 (V_j)\right] - 21 \sum_{j} 2\,k \mbox{\large Tr} \left[t_i^2(V_j) \large {\large \ln }\left(\frac{M_{V_j}}{M_U}\right)\right].
  \label{eq:threshold}
\end{eqnarray}
Here $t_{i}$ represents the generators of the super heavy particles under the $i^{\rm th}$ gauge group. Also the other factors are $k=\frac{1}{2}(1)$ for real (complex) scalars and $\kappa= \frac{1}{2}(1)$ is for Weyl (Dirac) fermions. Now, using the threshold effects in RGEs and after simplifications, one can derive the corrected unification mass scale as follows
\begin{eqnarray}
 &&{\large \ln}\left(\frac{M_U}{M_Z}\right) = 
 \frac{A_{I} {D_{W}}-B_{I} {D_{S}}}{B_{U} A_{I}-B_{I} A_{U}} + \frac{A_{I} {f^U_B}-B_{I} {f^U_A}}{B_{U} A_{I}-B_{I} A_{U}} \nonumber \\
 &&\hspace*{1.65cm}={\large \ln}\left(\frac{M_U}{M_Z}\right)_{1-{\rm loop}}+{\Delta}{\large \ln}\left(\frac{M_U}{M_Z}\right)_{\rm GUT-Th.}.
 \label{rel:threshold-a}
\end{eqnarray}
First term is the contribution from one-loop RGEs while the second term is for threshold corrections. The one-loop threshold corrections are contained in parameters like ${f^U_A}$ and ${f^U_B}$ which depend on ${\lambda}$'s as, ${f^U_A}= \left(8 {\lambda^U_{3C}}- 3 {\lambda^U_{2L}}- 5 {\lambda^U_{Y}}\right)/6$ and ${f^U_B}= \left(5 {\lambda^U_{2L}}- 5 {\lambda^U_{Y}}\right)/6$. This simplifies the corrections as
\begin{eqnarray}
 {\Delta} {\large \ln}\left(\frac{M_U}{M_Z}\right)
 &=&\frac{1}{700} \bigg[
 15 {\lambda^{U}_{Y}}(M_{U})- 13 {\lambda^{U}_{2L}}(M_{U})- 2 {\lambda^{U}_{3C}}(M_{U}) \bigg].
\end{eqnarray}
\noindent
{\bf For degenerate masses for super heavy fields:-}\,
For the estimation of threshold effects arising from super heavy particles, we assume that all the the super heavy gauge bosons have same mass but different from GUT symmetry breaking scale. The same assumption is also applicable to all other super heavy scalars and fermions. The estimated individual threshold corrections are
\begin{align} 
{\lambda^{U}_{3C}}(M_{U}) & = 5 -105 \eta_{V} + 70 \eta_{S} + 64 \eta_{F}, \nonumber\\
{\lambda^{U}_{2L}}(M_{U}) & = 6 -126 \eta_{V} + 68\eta_{S} + 48 \eta_{F},  \nonumber\\
{\lambda^{U}_Y}(M_{U}) & = 8 -168 \eta_{V} + \frac{308}{5} \eta_{S} + 64 \eta_{F}. 
 \end{align} 
Here, $
\eta_{S} = \mbox{ln}\frac{M_S}{M_U}$, $\eta_{F} = \mbox{ln}\frac{M_F}{M_U}$ and $\eta_{V} = \mbox{ln}\frac{M_{V}}{M_U}$. Using these values, the relation for unification mass scale with GUT threshold corrections is modified as,
\begin{align}
 {\Delta}{\large \ln}\left(\frac{M_U}{M_Z}\right) = \frac{1}{175} \bigg[
 8 - 168\,  \eta_{V} - 25\,  \eta_{S} + 52\,  \eta_{F}\bigg].
\end{align}
We have presented few benchmark points in Table. \ref{tab:Deg} for degenerate spectrum of super heavy particles and estimated the unification mass scale and proton lifetime including the threshold effects.
\begin{center}
\begin{table}[h]
\centering
\vspace{-2pt}
\begin{tabular}{||c|c|c|c|c|c|c|c|c|c|c||}
\hline \hline
$\pmb{\frac{M_{V}}{M_U}}$&$\pmb{\frac{M_{S}}{M_U}}$&$\pmb{\frac{M_F}{M_U}}$ & $\pmb{\lambda_{3C}}$ & $\pmb{\lambda_{2L}}$ & $\pmb{\lambda_{Y}}$ & $\pmb{M_U^{TH}}$ [GeV]& $\pmb{\tau_p}$ [yrs]   \\
\hline
 $\frac{1}{500}$ & $1.5$  & $\frac{1}{10}$  &$538.531$& $706.062$ & $929.631$ & $10^{15.5598}$ & $ 6.88124\times10^{33}$ \\
\hline
$\frac{1}{600}$ & $2.0$  & $\frac{1}{10}$  &$577.811$& $748.596$ & $977.981$ & $10^{15.6179}$ & $ 1.17508\times10^{34}$\\
\hline 

 $0.00123$ &  $ 2.03634$ & $0.10048$  &$610.978$& $787.971$ & $1029.96$ & $10^{15.7429}$ & $3.71592\times10^{34} $ \\   \hline
 
 $0.00107$ & $1.90066$   & $ 0.10013$  &$619.977$& $799.971$ & $1047.96$ & $10^{15.8025}$ & $ 6.43379\times10^{34}$ \\
\hline
 $ 0.00060$ & $1.6374$   & $0.0628$  &$640.976$& $840.969$ & $1106.96$ & $10^{15.9948}$  & $ 3.78152\times10^{35}$\\   \hline

 $0.00045$& $16.989$  & $0.0784$  &$849.969$& $1047.96$ & $1314.95$ & $10^{16.0017}$ & $ 4.02964\times10^{35}$ \\ 
 \hline
\hline 
\end{tabular}
\caption{Numerically estimated values for $M_U$ and ${\tau}_p$ by considering one-loop threshold effects.}
\label{tab:Deg}
\end{table} 
\end{center}
\noindent
{\bf For non-degenerate masses for super heavy vector bosons:-}\, Here, we assume all  super heavy color triplet and color singlet gauge bosons 
are non-degenerate but different from GUT symmetry breaking scale, while other super heavy scalars and fermions are degenerate. With new parameters like $\eta_{V_{C}}$ and $\eta_{V_{NC}}$ along with $\eta_{S}$ and $\eta_{F}$, the individual threshold corrections are estimated to  
\begin{align} 
{\lambda^{U}_{3C}}(M_{U}) & = 5 -21 \left(5 \eta_{V_{C}} + 0 \eta_{V_{NC}}\right) + 70 \eta_{S} + 64 \eta_{F}, \nonumber \\
{\lambda^{U}_{2L}}(M_{U}) & = 6 -21 \left(6 \eta_{V_{C}} + 0 \eta_{V_{NC}}\right) + 68\eta_{S} + 48 \eta_{F},  \nonumber\\
{\lambda^{U}_Y}(M_{U}) & = 8 -21 \left(\frac{34}{5} \eta_{V_{C}} + \frac{6}{5} \eta_{V_{NC}}\right) + \frac{308}{5} \eta_{S} + 64 \eta_{F}. 
\end{align} 
 where, 
$$\eta_{S} = \mbox{ln}\frac{M_S}{M_U}, \eta_{F} = \mbox{ln}\frac{M_F}{M_U}, \eta_{V_{C}} = \mbox{ln}\frac{M_{V_{C}}}{M_U}, 
\eta_{V_{NC}} = \mbox{ln}\frac{M_{V_{NC}}}{M_U}.$$ The notation $M_{VC}$ and $M_{V_{NC}}$ are the degenerate masses of the vector gauge bosons $V_4$ to $V_9$ and $V_1$ to $V_3$ respectively (shown in Table. \ref{tab:so10-superheavy}). Using these input values, GUT threshold corrected unification mass scale is given by
\begin{align}
{\Delta}{\large \ln}\left(\frac{M_U}{M_Z}\right) = \frac{1}{350} \bigg[
 16 - 147\,  \eta_{V_{C}} - 189\,  \eta_{V_{NC}} - 50\,  \eta_{S} + 104\,  \eta_{F}\bigg].
\end{align}
\begin{center}
\begin{table}[h]
\centering
\vspace{-2pt}
\begin{tabular}{||c|c|c|c|c|c|c|c|c|c|c||}
\hline \hline
$\pmb{\frac{M_{V_C}}{M_U}}$&$\pmb{\frac{M_{V_{NC}}}{M_U}}$ & $\pmb{\frac{M_{S}}{M_U}}$ & $\pmb{\frac{M_{F}}{M_U}}$& $\pmb{\lambda_{3C}}$ & $\pmb{\lambda_{2L}}$ & $\pmb{\lambda_{Y}}$ & $\pmb{M_U^{TH}}$ [GeV] & $\pmb{\tau_p} $ [yrs]   \\
\hline
$ \frac{1}{400} $ &  $\frac{1}{100} $ & $2.0 $ &  $0.3 $ &  $617.914 $ & $765.08 $ & $962.059 $ & $10^{15.287}$ &$5.57785\times10^{32} $ \\
\hline
$\frac{1}{2500} $ &  $\frac{1}{100} $ & $2.5 $ &  $0.5 $ & $846.273 $ & $1020.83 $ & $1253.36 $ & $10^{15.6519}$ & $1.60719\times10^{34} $\\ 
\hline
$\frac{1}{3000}$ &  $\frac{1}{300} $ & $1.5 $ &  $0.25 $ & $ 785.299 $ & $ 975.796$ & $ 1231.25$ & $10^{15.885}$ & $1.37552\times10^{35} $\\  
\hline
$\frac{1}{4000} $ &  $\frac{1}{400} $ & $ 0.5$ &  $ 0.2$ & $ 724.324 $ & $926.629 $ & $1197.63 $ & $10^{16.044}$   & $5.94931\times10^{35} $\\  \hline
$\frac{1}{5000} $ &  $\frac{1}{400} $ & $0.15 $ &  $ 0.2$ &  $663.479 $ & $872.877 $ & $1155.33 $ & $10^{16.16}$ & $1.73168\times10^{36} $\\  
\hline
$\frac{1}{2000}$ &  $\frac{1}{500} $ & $\frac{1}{2500} $ &  $0.1 $ & $108.042 $ & $321.143 $ & $ 620.668$ & $10^{16.323}$ & $7.77082\times10^{36} $\\  
\hline 
\hline 
\end{tabular}
\caption{Numerically estimated values of $M_U$ and ${\tau}_p$ by including one-loop threshold effects by considering non-degenerate masses for super heavy vector bosons.}
\label{tab:NonDeg}
\end{table} 
\end{center}
In Table \ref{tab:NonDeg}, we have presented various mass values for gauge bosons, scalars and fermions and estimated the threshold corrected unification mass scale and the corresponding proton lifetime.
The evolution of gauge coupling constants including one-loop threshold corrections is shown in Fig. \ref{plot:unifn_threshold} by considering the benchmark given the last row of Table. \ref{tab:NonDeg}. The corresponding corrected unification mass scale and proton lifetime consistent with Super-Kamiokande \cite{Miura:2016krn}and Hyper-Kamiokande \cite{Abe:2011ts,Yokoyama:2017mnt}, are
\begin{align}
M_U^{TH}=10^{16.323}\,\mbox{GeV},  \quad \tau_p = 7.77082\times10^{36}\mbox{yrs}.
\end{align}
\section{Discussion on fermion masses and mixing} 
It has been pointed in Witten's work~\cite{Witten:1979nr} that the minimal non-supersymmetric $SO(10)$ model with only $10_H$ (containing SM Higgs) and $16_F$ (accommodating SM fermions plus right-handed neutrinos) predicts $m_d \simeq m_e$, $m_s \simeq m_\mu$, which are ruled out from experiments. In this scenario, the neutrinos have only Dirac masses proportional to up-type quark masses which disagree with the current neutrino oscillation data. Moreover, there is a possibility to have small Majorana masses for right-handed neutrinos via two-loop effects using $16_H$ on top of $10_H$. Thus, the failure of $SO(10)$ model to account correct fermion masses and mixing motivates us to explore all possible non-minimal scenarios. 

The correct fermion masses and mixing can be addressed in non-supersymmetric $SO(10)$ theory with the inclusion of extra $10_H$~\cite{Bajc:2005zf} while $126_H$ was introduced for breaking of Pati-Salam symmetry or left-right symmetry as an intermediate symmetry between $SO(10)$ theory and SM \cite{PhysRevD.31.2316,Anastaze:1983zk,Yasue:1980qj,Yasue:1980fy,Bertolini:2009qj,Bertolini:2009es,Bajc:2005zf,Altarelli_2013,Dueck_2013,Joshipura:2011nn,Babu:2016bmy,Ohlsson:2018qpt}. In our present work, we adopt two $10_H$ and one $126_H$ representation along with $16_F$ and $45_F$ in the fermion sector in order to explain fermion masses and mixing, dark matter and flavor physics anomalies. The SM Higgs can be admixture of two Higgs doublets contained in two $10_H$'s and also from $126_H$. We assume that Pati-Salam symmetry of left-right symmetry is broken at GUT breaking scale and that is the reason why we are expecting the right-handed neutrinos as well as scalar triplets can have their masses around $10^{13}$ GeV close to $M_U$. With this, we have not considered its effects while numerically examining the renormalization evolution equation for gauge couplings.

The $SO(10)$ invariant Yukawa interactions are
\begin{eqnarray}
\mathcal{L}_{Y} = 16^i_F \left( Y^{ij}_{10}\, 10_H + Y^{ij}_{126} \overline{126_H}  \right) 16^j_F + M_{45} 45_F 45_F +  {\rm h.c}.,
\end{eqnarray}
where $Y_{10}$ and $Y_{126}$ are complex symmetric matrices. Here, $SO(10)$ or equivalent Pati-Salam symmetry ($SU(4)_C \times SU(2)_L \times SU(2)_R$) is broken down at the predicted unification mass scale $10^{13.27}$ GeV. As a result, since the right-handed symmetry breaking scale is close to $M_U$, the right-handed neutrinos as well scalar triplets have their masses around that scale, which can generate neutrino masses and mixing via type-I and type-II seesaw mechanisms respectively. 

Using two real representations $10_{H_1}$ and $10_{H_2}$, an equivalent complex $10_H$ can be constructed as $10_H = 10_{H_1}+i\,10_{H_2}$,  without effecting the evolution of gauge couplings. Additionally, we introduce a global Pecci-Quinn symmetry forbidding the Yukawa couplings involving $10^*_H$ \cite{PhysRevLett.38.1440,PhysRevD.16.1791}. The $U(1)_{PQ}$ transformation of the relevant $SO(10)$ representations are as follows,
\begin{eqnarray}
&&16_F \to e^{i\,\alpha} 16_F, \quad  45_F \to  45_F,\, \nonumber \\
&&10_H \to e^{-2i\,\alpha} 10_H, \quad  \overline{126_H} \to e^{-2i\,\alpha} \overline{126_H}\,.
\end{eqnarray}
As a result of this Pecci-Quinn symmetry, we have two separate vevs (vacuum expectation values) $v_{u,d}^{10} \subset 10_H$ and one Yukawa coupling $Y_{10}$. The other advantage of Pecci-Quinn symmetry\cite{Boucenna:2018wjc} is to solve strong CP problem and provide axion dark matter \cite{Mohapatra:1982tc,Marsh:2015xka,PhysRevLett.53.1292,Jeong:2022kdr,PhysRevLett.40.223,PhysRevLett.40.279,Ringwald:2012hr,Graham:2015ouw,Arias:2012az,Langacker:1986rj}.

The Yukawa terms for fermion masses and mixing, relevant at the PS symmetry breaking scale are given by
\begin{eqnarray}
\mathcal{L}^{\rm PS}_{\rm Yuk} = Y^{ij}_{10_F} F^{i^T}_{L} \Phi F^{j}_{R} +\widetilde{Y}^{ij}_{10_F} F^{i^T}_{L} \widetilde{\Phi} F^{j}_{R}
+ Y^{ij}_{126_F} F^{i^T}_{L} \overline{\xi} F^{j}_{R} + 
   Y^{ij}_{126_R} F^{i^T}_{R} \overline{\Delta}_R F^{j}_{R} +  M_{\Sigma} \Sigma_F \Sigma_F +  {\rm h.c}.\nn\\ 
\end{eqnarray}
The details of mapping of Yukawa couplings at $SO(10)$ and Pati-Salam symmetry can be understood by solving RGEs and interested reader may refer to ~\cite{Ohlsson:2021lro}. Here we consider the vevs $v_1$ and $v_2$ from the PS multiplet $(1,2,2)$ of $10_H$. The primary role of these vevs (of the order of EW scale) is to generate fermion masses and help in breaking of $SU(3)_C \times \SU(2)_L \times U(1)_Y $ down to $SU(3)_C \times U(1)_{\rm em}$. The other vevs are relevant for correcting bad mass relations in fermion masses are $\langle \xi \rangle \approx v_{\xi_1}, v_{\xi_2} \subset 126_{H}$ in the MeV scale. As pointed out in ref~\cite{Babu:1992ia}, these small induced vevs of Pati-Salam multiplet $(15_{4C}, 2_{2L}, 2_{2R})$ are coming from the important scalar interaction term,
$$
V=\lambda_{\xi}\,M^{\prime}\,{210}_H\,{126}_H^{\dagger}\,{10}_H \supset \lambda_{\xi}\,M^{\prime}\,{(15,2,2)}_{126}\,
{(15,1,1)}_{210}\,{(1,2,2)}_{10}\, ,$$
This provides the induced vev $v_{\xi}$ of the neutral component of $\xi(15,2,2)$ as,
$$v_{\xi} = \lambda_{\xi}\,M^{\prime}\,M_C\,v_{ew}/M_{\xi}^2\, , $$
where, $v_{ew} = \sqrt{v_1^2 + v_2^2}$. Using these vevs and Yukawa couplings, the fermion masses at electroweak scale can be  expressed in terms of Yukawa couplings defined at Pati-Salam scale and various vevs arising from $10_H$ and $126_H$ are given by
\begin{eqnarray}
& & M_u \equiv v_u Y_u = v_{1} Y_{10_F}^u + v_{\xi_1} Y_{126_F}^{u}\, ,\quad  M_d \equiv v_d Y_d = v_{2} Y_{10_F}^d + v_{\xi_2} Y_{126_F}^{d}\, , \nonumber \\ 
& & M_e \equiv v^{10}_d Y_{e} = v_{2} Y_{10_F}^e + v_{\xi_2} Y_{126_F}^{e}\, ,  \quad  M^D_\nu \equiv v_u Y_\nu = v_{1} Y_{10_F}^\nu + v_{\xi_1} Y_{126_F}^{\nu}\, , \nonumber \\ 
& & M_R    = v_R Y^{\nu}_{126_R}\, , \quad \big[M_L  = v_L Y^{\nu}_{126_L} \, \mbox{for $G_{224D}$} \big]. 
\label{eq:massrel-one10}
\end{eqnarray}
Here, $M_u$ ($M_d$) denotes the mass matrix for up (down)-type quarks whereas $M_e$ represents the mass matrix of the charged leptons. Also $M^D_\nu$ is Dirac neutrino mass matrix, while $M_L$  and $M_R$ stand for Majorana mass matrix for light left-handed and heavy right-handed neutrinos respectively. Applying appropriate boundary condition at Pati-Salam symmetry breaking scale, the simplified fermion mass matrices become~\cite{Babu:1992ia},
\begin{eqnarray}
& & M_u =  H v_1 + F v_{\xi_1} \, , \quad M_d =  H v_2 + F v_{\xi_2}\nonumber \\ 
& & M^D_\nu = H v_1 -3 F v_{\xi_1} \, , \quad M_e = H v_2 -3  F v_{\xi_2} \nonumber \\
& & M_R    = F v_R
\label{eq:massrel-one10}
\end{eqnarray}
Here, $H$ and $F$ are Yukawa coupling matrices derived in terms Yukawa coupling matrices defined at Pati-Salam symmetry. Let us consider a basis where 
$H$ is real and diagonal. Also define two more parameters; ratio between two Higgs doublet VEVs of $10_H$ i.e, $r_1 = v_2/v_1$ and ratio between two Higgs doublet VEVs of $126_H$ i.e, $r_2=v_{\xi_2}/v_{\xi_1}$. As a result, there are total 13 parameters excluding the VEVs $v_R$ (or $v_L$) present in the fermion mass fitting: 
3 diagonal elements of matrix $H$, 6 elements of symmetric matrix F, 2 ratios of VEVs  $r_1, r_2$ and two physical phases $\alpha$ and $\beta$ used in the VEVs. These 13 parameters have been utilised to explain the 13 observables in the charged fermion masses: 9 fermion masses, 3 quark mixing angles and one CP-phase. Also, the resulting Dirac neutrino mass matrix can be expressed in terms of $v_R$ and other input model parameters. So, one can rewrite the simplified fermion mass relations in terms of these ratios of different VEVs as follows~\cite{Babu:1992ia},
\begin{eqnarray}
&&M_e = \frac{4 r_1 r_2}{r_2-r_1} M_u - \frac{r_1 + 3 r_2}{r_2-r_1} M_d\, , 
 \nonumber \\
&& M^D_\nu = \frac{3 r_1 +r_2}{r_2-r_1} M_u - \frac{4}{r_2-r_1} M_d\, \nonumber \\
&& M_R = \frac{1}{R} \frac{r_1}{r_1-r_2} M_u - \frac{1}{R} \frac{1}{r_1-r_2} M_d\,,
\end{eqnarray}
where $R = v_1/v_R$. We can consider a basis where $M_u$ is already diagonal with masses as $M_u = \mbox{Diag}\big(m_u, m_c, m_t \big)$. In this choice of basis, the down-type quark mass matrix can be diagonalised by $\hat{M}_d \simeq V^\dagger_{\rm CKM} M_d V_{\rm CKM} = \mbox{Diag}\big(m_d, m_s, m_b \big)$ where $V_{\rm CKM}$ is the usual CKM mixing matrix. It is to be noted that the charged lepton mass matrix can now be fully determined in terms of physical observables of quark sector and two parameters related to ratios of various VEVs i.e, $r_1$ and $r_2$.

Let us consider that the Dirac neutrino mass matrix is approximated to be up-quark mass matrix in the present scenario with the high scale intermediate symmetry as Pati-Salam. Using the seesaw approximation with the mass hierarchy $M_R \gg M_\nu^D \gg M_L$, the resulting light neutrino mass formula via type-I seesaw with the PS symmetry without D-parity as the only intermediate symmetry or type-I+II within D-parity conserving PS symmetry, seesaw contributions are as follows
\begin{equation}
M_\nu=- M^D_\nu M^{-1}_R M^D_\nu\, \big( + M_L \quad \mbox{for}\, G_{422D} \big)\,.
\label{eq:seesaw-numass}
\end{equation}
For typical value of $M_R \sim 10^{13.27}$ GeV, $M^D_\nu \sim 100$ GeV, we obtain sub-eV mass for light neutrinos. The out-of-equilibrium decays of right-handed neutrinos can provide the observed baryon asymmetry of the Universe via type-I leptogenesis. We skip the details of fermion mass fitting and its implications to matter-antimatter asymmetry of the universe which can be looked up in recent works~\cite{Babu:1992ia,Mummidi:2021anm}.

\section{Addressing flavor anomalies with scalar leptoquark $R_2$}
It has been already examined that inclusion of TeV scale SLQ and a fermion triplet DM candidate leads to successful  unification of the gauge coupling, when embedded in a non-supersymmetric $SO(10)$ GUT. The presence of TeV scale SLQ arising from GUT framework has interesting low-energy phenomenology like explaining flavor anomalies, muon $g-2$, collider studies etc \cite{Babu:2020hun}. However, in the present work, we stick with discussions of phenomenological implications of SLQ to recent flavor anomalies  in semileptonic $B$ decays.  In recent times, several intriguing deviations  at  $(2-4)\sigma$ significance level, have been realized by the three pioneering experiments: Babar \cite{Lees:2012xj, Lees:2013uzd}, Belle \cite{Huschle:2015rga, Hirose:2016wfn,Abdesselam:2019wac,Belle:2019gij, Abdesselam:2019lab}  and LHCb \cite{Aaij:2013aln, Aaij:2013qta, Aaij:2014pli, Aaij:2014ora,  Aaij:2015yra, Aaij:2015esa, Aaij:2017vbb,  Aaij:2017uff, Aaij:2017tyk, Aaij:2019wad},  in the form of lepton flavour universality (LFU) violation associated with the charged-current (CC)  and neutral-current (NC)   transitions  in semileptonic $B$ decays. These discrepancies can't be accommodated in the SM and are generally interpreted as smoking-gun signals  of NP contributions. The discrepancies in the CC sector are usually attributed to the presence of new physics in $ b \to   c \tau \bar \nu_\tau$ transition, whereas in the NC sector to $b \to s \mu \mu$ process. It has been shown in the literature that various leptoquark scenarios can successfully address these anomalies. Here, we will show that $R_2(3,2,7/6)$ leptoquark present in our model  can successfully explain these discrepancies.

The generalized effective Hamiltonian accountable for the charged-current $b \to c \tau \bar{\nu}_\ell$  transitions is given as  \cite{Tanaka:2012nw}
\bea \label{ham-bclnu}
\mathcal{H}_{\rm eff}^{\rm CC} \ = \ \frac{4G_F}{\sqrt{2}} V_{cb} \Big [ \left(\delta_{\ell\tau} + C_{V_1}^\ell \right) \mathcal{O}_{V_1}^\ell + C_{V_2}^\ell \mathcal{O}_{V_2}^\ell +  C_{S_1}^\ell \mathcal{O}_{S_1}^\ell +  C_{S_2}^\ell \mathcal{O}_{S_2}^\ell+ C_{T}^\ell \mathcal{O}_{T}^\ell \Big ],
\eea
where $G_F$ and  $V_{cb}$ represent the Fermi constant and the Cabibbo-Kobayashi-Maskawa (CKM) matrix element respectively.  $C_X^\ell$ are the new Wilson coefficients, with $X=V_{1,2}, S_{1,2}, T$, which  can  arise only  when  NP prevails. The  corresponding four-fermion operators ${\cal O}_X^\ell$ can be expressed as 
\bea
&&\mathcal{O}_{V_1}^\ell \ = \ \left(\bar{c}_L \gamma^\mu b_L \right) \left(\bar{\tau}_L \gamma_\mu \nu_{\ell L} \right), \qquad 
\mathcal{O}_{V_2}^\ell \ = \ \left(\bar{c}_R \gamma^\mu b_R \right) \left(\bar{\tau}_L \gamma_\mu \nu_{\ell L} \right), \nonumber \\
&&\mathcal{O}_{S_1}^\ell \ = \ \left(\bar{c}_L  b_R \right) \left(\bar{\tau}_R \nu_{\ell L} \right), \qquad \qquad 
\mathcal{O}_{S_2}^\ell \ = \ \left(\bar{c}_R b_L \right) \left(\bar{\tau}_R \nu_{\ell L} \right), \nn \\
&&\mathcal{O}_{T}^\ell \ = \ \left(\bar {c}_R \sigma^{\mu \nu}  b_L \right) \left(\bar{\tau}_R \sigma_{\mu \nu} \nu_{\ell L} \right)\,,
\eea
where $f_{L(R)} = P_{L(R)}f$  with $P_{L(R)}=(1\mp \gamma_5)/2$, represent the chiral fermion fields $f$. 

The effective Hamiltonian delineating  the NC transitions $b \to s \ell^+ \ell^-$  is given  as  \cite{Misiak:1992bc, Buras:1994dj}
\bea
{\cal H}_{\rm eff}^{\rm NC} \ = \ - \frac{ 4 G_F}{\sqrt 2} V_{tb} V_{ts}^* \Bigg[\sum_{i=1}^6 C_i(\mu) \mathcal{O}_i +\sum_{i=7,9,10,S, P} \Big ( C_i(\mu) \mathcal{O}_i
+ C_i'(\mu) \mathcal{O}_i' \Big )
\Bigg]\;,\label{ham-bsll}
\eea
where  $V_{tb}V_{ts}^*$ represents the product of CKM matrix elements,  $C_{i}$'s denote the Wilson coefficients   and $\mathcal{O}_i$'s are the   four-fermion  operators expressed as:  
\bea
\mathcal{O}_7^{(\prime)} & \ = \ &\frac{\alpha_{\rm em}}{4 \pi} \bigg[\bar s \sigma_{\mu \nu}
\big \{m_s P_{L(R)} + m_b P_{R(L)} \big \} b\bigg] F^{\mu \nu}, \nonumber \\
\mathcal{O}_9^{(\prime)}& \ = \ & \frac{\alpha_{\rm em}}{4 \pi} \big(\bar s \gamma^\mu P_{L(R)} b\big)(\bar \ell \gamma_\mu \ell)\;, \qquad \mathcal{O}_{10}^{(\prime)} \ = \ \frac{\alpha_{\rm em}}{4 \pi} \big(\bar s \gamma^\mu 
P_{L(R)} b \big)(\bar \ell \gamma_\mu \gamma_5 \ell)\; , \nonumber \\
\mathcal{O}_S^{(\prime)}& \ = \ & \frac{\alpha_{\rm em}}{4 \pi} \big (\bar s  P_{L(R)} b \big )(\bar \ell  \ell)\;,\qquad \qquad \mathcal{O}_{P}^{(\prime)} \ = \ \frac{\alpha_{\rm em}}{4 \pi} \big (\bar s  
P_{L(R)} b\big )(\bar \ell  \gamma_5 \ell) \, .
\eea
The  primed   as well as scalar/pseudoscalar operators are absent in the SM and can be generated only in beyond the SM scenarios.
\subsection{New contributions with scalar leptoquark} \label{sec:model}
\begin{figure}[t!]
	\centering
	\hspace*{-0.5cm}
	\includegraphics[width=0.9\textwidth]{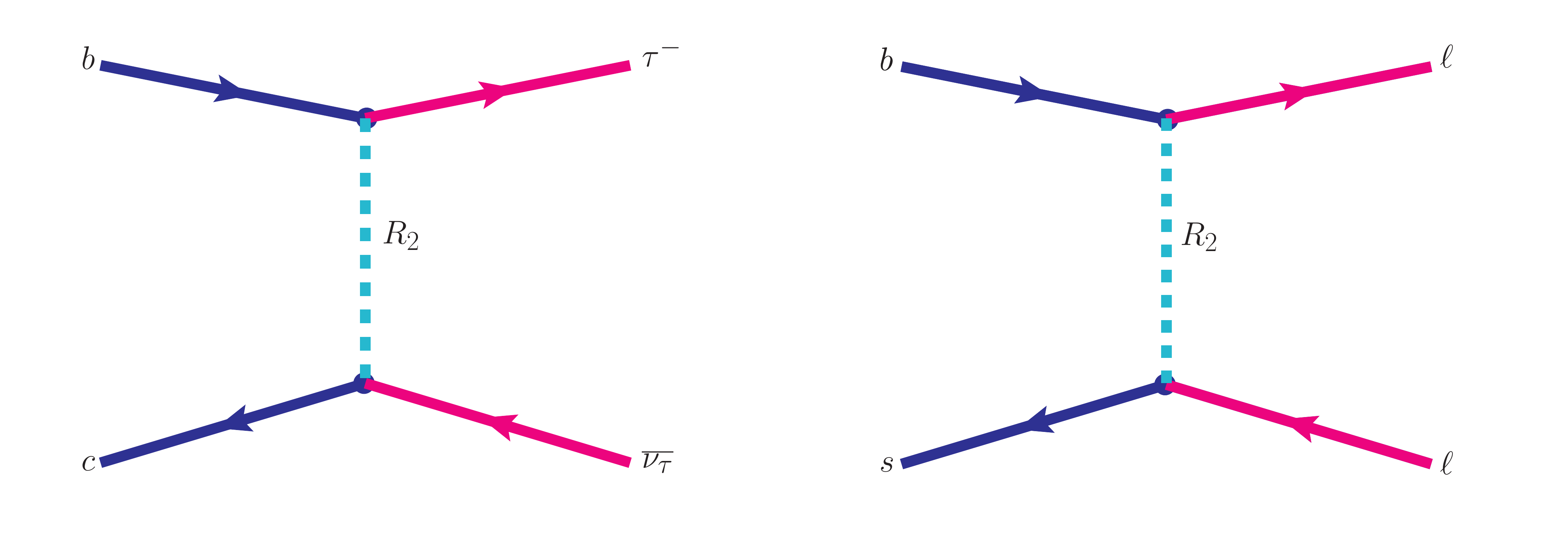}
	\caption{Feynman diagram for the CC transition  $b \to c \tau^- \bar \nu_\tau$  (left panel) and NC process $b \to s \ell^+ \ell^-$ (right panel)  induced by the scalar leptoquark $R_2^{(2/3)}$. }
	\label{fig:Fyn-LFC}
\end{figure}

In the context of the present model, the flavour sector will be sensitive to the presence of the SLQ $R_2(3,2,7/6)$, which  can provide additional contributions to the CC mediated  $b \to c \ell \bar \nu$  as well as NC $b \to s \ell^- \ell^+$ processes and can elucidate the observed  data reasonably well.  
The SLQ couples  simultaneously to quark and lepton fields through flavor dependent Yukawa couplings and the corresponding  interaction Lagrangian  can be written as \cite{ Iguro:2018vqb, Sakaki:2013bfa},  
\begin{eqnarray}
{\cal L}_{\rm int}=\lambda_R^{ij} \overline Q_{Li} \ell_{Rj} R_2 - \lambda_L^{ij} \overline u_{Ri} R_2 i \tau_2 L_{Lj} +{\rm h.c.},\label{Int-L}
\end{eqnarray}
where the couplings $\lambda_{L,R}$ are in general $3 \times 3$ complex matrices, $R_2 = \left(R^{(5/3)}_2~ R^{(2/3)}_2 \right)^T$, $Q_L (L_L)$ represents the left-handed quark (lepton) doublet, $u_R (\ell_R)$ is the right-handed singlet up-type quark (charged lepton) and  the generation indices are characterized by $i,j$.  The interaction Lagrangian (\ref{Int-L}) in the mass basis  can be obtained after the expanding the $SU(2)$ indices as \cite{ Iguro:2018vqb}
\begin{eqnarray}
{\cal L}_{\rm int}&=&(V_{\rm CKM}\lambda_R)^{ij} \overline u_{Li} \ell_{Rj} R_2^{(5/3)} +\lambda_R^{ij} \overline d_{Li} \ell_{Rj} R_2^{(2/3)}\nn\\
&+& \lambda_L^{ij} \overline u_{Ri} \nu_{Lj} R_2^{(2/3)}  -\lambda_L^{ij} \overline u_{Ri} \ell_{Lj} R_2^{(5/3)} +{\rm h.c.}.
\label{Int-L1}
\end{eqnarray}
Here the superscripts on $R_2$ specify its electric charge and the mass bases for quark doublets are considered as $ ( (V_{\rm CKM}^\dagger u_L)^i, d_L^i )^T$ while for lepton doublets as  $(\nu_L^i, \ell_L^i)^T$, neglecting the mixing  in the lepton sector.  Thus,   from eqn. (\ref{Int-L1}), one can notice  that the exchange of $R_2^{(2/3)}$ can induce new contribution to both $b \to c \tau \bar \nu_\tau$  as well as $b \to s \mu^- \mu^+$ transitions at tree-level as shown in Fig. \ref{fig:Fyn-LFC}. 

For  $b \to c \tau \bar \nu_\tau$ it generates additional scalar as well as tensor interactions at the LQ mass scale $(\mu=m_{\rm LQ})$ as: 
\bea
C_{S_2}^{\rm NP}(m_{\rm LQ})=4 C_T^{\rm NP} (m_{\rm LQ}) = \frac{1}{4 \sqrt 2 G_F V_{cb}} \frac{\lambda_L^{23} (\lambda_R^{33})^*}{m_{\rm LQ}^2}\;,\label{SLQ}
\eea
where $m_{\rm LQ}$ represents the  leptoquark mass, and we consider a typical TeV scale SLQ  in our analysis. It should be noted that the new Wilson coefficients as shown  in eqn. (\ref{SLQ}) rely on the LQ mass scale $\mu(m_{\rm LQ})$, and hence, it is  essential to evolve their values from the $m_{\rm LQ}$ scale to the  $b$-quark mass scale $\mu=m_b$ through  the renormalization-group  equation (RGE), which are expressed as \cite{Blanke:2018yud, Gonzalez-Alonso:2017iyc} 
\bea
\begin{pmatrix}
{C_{S_2}^{\rm NP}}(m_b)\\
{C_T^{\rm NP}}(m_b)
\end{pmatrix}=  \begin{pmatrix} 1.752 & -0.287\\
-0.004 & 0.842  \end{pmatrix}        
\begin{pmatrix}{C_{S_2}^{\rm NP}}(m_{\rm  LQ})\\
{C_T^{\rm NP}}(m_{\rm  LQ})
\end{pmatrix}. \label{SLQ1}
\eea

Similarly, after performing the Fierz transformation,   the  new contribution to the  $b \to s \mu^+ \mu^- $ process  can be   obtained from Eq. (\ref{Int-L1}) as,
\bea
{\cal H}_{\rm LQ}&=& \frac{\lambda_R^{32} {(\lambda_R^{22 })}^* }{8 m_{\rm LQ}^2} \left ( \bar s \gamma^\mu (1-\gamma_5)b \right ) \left (
\bar \mu \gamma_\mu(1+\gamma_5) \mu \right )\equiv \frac{\lambda_R^{32} {(\lambda_R^{22 })}^*}{4 m_{\rm LQ}^2} \Big (O_9 +O_{10} \Big).\label{LQ-b2s}
\eea
 Thus, comparing (\ref{LQ-b2s}) with (\ref{ham-bsll}), one can obtain the new Wilson coefficients as 
\bea
C_9^{\rm NP} = C_{10}^{\rm NP} = - \frac{ \pi}{2 \sqrt 2 G_F \alpha_{\rm em} V_{tb} V_{ts}^* }\frac{\lambda_R^{32}{ (\lambda_R^{22 })}^*}{
m_{\rm LQ}^2}\;.\label{c10np}
\eea
After delineating the additional contributions to the Wilson coefficients for the $b \to c \tau \bar \nu$ and $b \to s \mu^+ \mu^-$ transitions,  we now proceed to constrain the new parameters. We perform a global-fit using all the relevant experimental observables to constrain these new couplings.  The  list of the observables are provided in the following subsection. 
\subsection{List of observables used in global-fit} \label{sec:obs}
In this analysis, we incorporate the following    observables  for constraining  the new couplings.  
\begin{enumerate}

\item {\bf Observables associated with $\boldsymbol{b \to s \mu^+ \mu^-}$ transitions}: 
\begin{itemize}
\item {$\boldsymbol{R_K}$ {\bf and} $\boldsymbol{R_{K^*}}$ :}
The LFU violating observables $R_K$ and $R_{K^*}$,  expressed as 
\begin{align} \label{Eqn:RK}
R_{K} \ & = \ \frac{{\rm BR}(B^+ \to K^{+} \mu^+ \mu^-)}{{\rm BR}(B^+ \to K^{+} e^+ e^-)} \, , \qquad 
R_{K^{*}} \  = \ \frac{{\rm BR}(B^0 \to  K^{*0} \mu^+ \mu^-)}{{\rm BR}( B^0 \to  K^{*0} e^+ e^-)} \, .
\end{align}
The recently updated  values of  $R_K$ \cite{LHCb:2021trn} and $R_{K^*}$  \cite{Aaij:2017vbb}, by LHCb experiment in the low $q^2$ bins are given as:
\bea \label{Eqn:RK-Exp-new}
R_K^{\rm LHCb} =  0.846^{+0.044}_{-0.041}\,  ,\qquad q^2\in [1.1, 6]~{\rm GeV}^2 \, ,
\eea
\bea
R_{K^*}^{\rm LHCb}&  =  & \begin{cases}0.660^{+0.110}_{-0.070}\pm 0.024 \qquad q^2\in [0.045, 1.1]~{\rm GeV}^2 \, , \\ 
0.685^{+0.113}_{-0.069}\pm 0.047 \qquad q^2\in [1.1,6.0]~{\rm GeV}^2 \, .
\end{cases}
\eea 
In addition to  the LHCb results, the Belle experiment also has recently reported new measurements on $R_K$~\cite{Abdesselam:2019lab} and 
$R_{K^*}$~\cite{Abdesselam:2019wac} in several other bins.
However, as  the  Belle  results have comparatively larger uncertainties,  we do not consider them in our fit for constraining the  new parameters.

\item {$\boldsymbol{B_s \to \mu^+ \mu^-}$ :} 

The  current average value on the branching franction of  $B_s \to \mu^+ \mu^-$ process  from the combined results of ATLAS, CMS and LHCb is ~\cite{ATLAS:2020acx}: 
\begin{align}
{\rm BR}(B_s^0\to \mu^+\mu^-) \ = \ \left(2.69^{+0.37}_{-0.35}\right)\times 10^{-9} \, ,
\end{align}
which has  $2.4\sigma$ deviation from its SM prediction~\cite{Bobeth:2013uxa} 
\begin{align}
{\rm BR}(B_s^0\to \mu^+\mu^-)^{\rm SM} \ = \ \left(3.65\pm 0.23\right)\times 10^{-9} \, .
\end{align}

\item {\bf $\boldsymbol{B\to K^*\mu\mu}$ and $\boldsymbol{B_s\to \phi \mu\mu}$ processes:} 

$\bullet$ 
 We consider the  following set of angular observables from $ B^0 \to K^{*0}  \mu^+ \mu^-$  process: the form factor independent optimized observables $P_{1,2,3}, P_{4,5,6,8}^\prime$,  the longitudinal polarization fraction $(F_L)$ and the forward-backward asymmetry  $(A_{FB})$   in the   following $q^2$ bins (in ${\rm GeV}^2$): $[0.1, 0.98], ~[1.1, 2],~[2,3], ~[3, 4], ~[4, 5], ~[5, 6]$ and  $[1, 6]$ \cite{LHCb:2015svh}.\\

$\bullet$  For  $B_s \to\phi \mu^+ \mu^-$ mode, we take into account the longitudinal polarization asymmetry $(F_L)$ and CP averaged  observables ($S_{3,4,7}, A_{5,6,8,9}$)   in the following three $q^2$ bins  (in ${\rm GeV}^2$): $[0.1, 2],~ [2, 5]$, and $[1, 6]$ \cite{Aaij:2015esa}.

\end{itemize}

\item {\large $\boldsymbol{b \to c \tau \bar \nu_\tau}$}: For the CC transitions $b \to c \tau \nu$, we incorporate the  following observables. 
\begin{itemize}
\item {$\boldsymbol{R_D}$ {\rm and} $\boldsymbol{R_{D^{*}}}$ :} The lepton non-universality observables $R_D$ and $R_{D^*}$,  defined as 
\begin{align}
R_{D^{(*)}} \ = \ \frac{{\rm BR}(B \to D^{(*)}\tau \bar \nu_\tau)}{{\rm BR}(B \to D^{(*)}\ell \bar \nu_\ell)} \, ,
\end{align}
with $\ell=e,\mu$. 
These observables  are measured by  BaBar \cite{Lees:2012xj,Lees:2013uzd} and Belle~\cite{Huschle:2015rga,Hirose:2016wfn,Abdesselam:2016cgx} whereas only  $R_{D^{*}}$ has been measured by LHCb~\cite{Aaij:2015yra,Aaij:2017uff}. The present world-average values of these ratios obtained by  incorporating the data from all these measurements are~\cite{HFLAV:2019otj}:
\bea
R_{D}^{\rm exp} =  0.34\pm 0.027\pm 0.013\,,~~~~~
R_{D^*}^{\rm exp}=  0.295\pm 0.011\pm 0.008\,,
\eea
exhibit   $3.08\sigma$ discrepancy with the corresponding SM results \cite{Fajfer:2012vx, Fajfer:2012jt}
\bea
R_{D}^{\rm SM}  \ = \ 0.299\pm 0.003\,,~~~~~~
R_{D^*}^{\rm SM}  \ = \ 0.258\pm 0.005\,.
\eea

\item {$\boldsymbol{R_{J/\psi}}$ :}
Analogously,  in the  measurement  of $R_{J/\psi}$~\cite{Aaij:2017tyk}
\bea
R_{J/\psi}^{\rm exp} \ = \ \frac{{\rm BR}(B \to J/\psi\tau \bar \nu_\tau)}{{\rm BR}(B \to J/\psi \ell \bar \nu_\ell)} \ = \ 0.71\pm 0.17\pm 0.184\,,
\eea
 a discrepancy of about $1.7 \sigma$  has been observed with the  corresponding SM prediction~\cite{Ivanov:2005fd, Wen-Fei:2013uea,  Watanabe:2017mip}
\bea
R_{J/\psi}^{\rm SM} \ = \ 0.289\pm 0.01\, . 
\eea

\item {$\boldsymbol{B_c^+\to \tau^+\nu_\tau}$ :} This leptonic decay process has not been observed so far,  however, indirect constraints on ${\rm BR}(B_c^+\to \tau^+\nu_\tau)\lesssim 30\%$ has been enforced  using the lifetime of $B_c$~\cite{Alonso:2016oyd, Li:2016vvp, Celis:2016azn}.
\end{itemize}

\end{enumerate}
%
For the numerical estimation of the  SM results of the above-mentioned observables, we use  the   masses of various particles  and the lifetime of $B_q$ mesons  from PDG~\cite{Zyla:2020zbs}. The SM result for ${\rm BR}(B_s \to \mu^+ \mu^-)$  is taken from Ref.~\cite{Bobeth:2013uxa}.  For evaluating the  $B \to K$ transition form factors we  use the light cone sum rule (LCSR) approach \cite{Ball:2004ye} and for  $B_{(s)} \to K^*(\phi)$  transitions, we use the form factors from Refs.~\cite{Ball:2004rg, Beneke:2004dp}. The expressions for the decay rates for $B \to D^{(*)} \ell \nu$ and $B_c  \to J/\psi \ell \nu$ are taken from \cite{Sakaki:2013bfa}. 
The form factors used for processes involving $b \to c$ transitions are as:  $B \to D$  \cite{MILC:2015uhg}, $B \to D*$ \cite{FermilabLattice:2014ysv, HeavyFlavorAveragingGroupHFAG:2014ebo} and for $B_c \to J/\psi$ \cite{Watanabe:2017mip}. 
The  $B_c$ meson decay constant   is considered as $f_{B_c}=489$ MeV \cite{Chiu:2007km} for computing  ${\rm BR}(B_c \to \tau \nu_\tau)$ and its expression is taken from \cite{Watanabe:2017mip}.

\subsection{Numerical fits of model parameters} \label{sec:fit}

\begin{figure*}[t!]
\centering
{\includegraphics[width=0.4\textwidth]{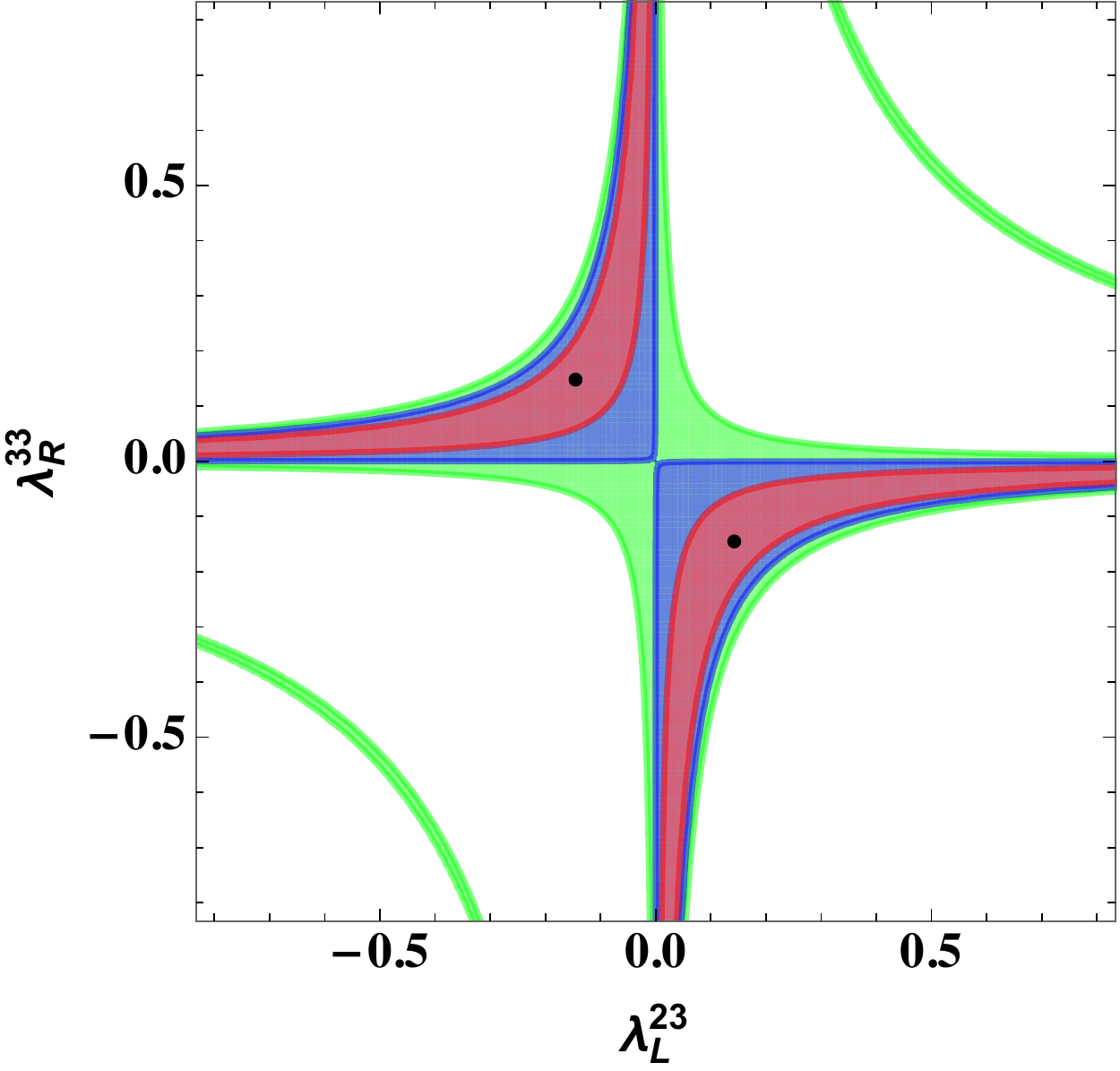}}
\quad \quad
{\includegraphics[width=0.4\textwidth]{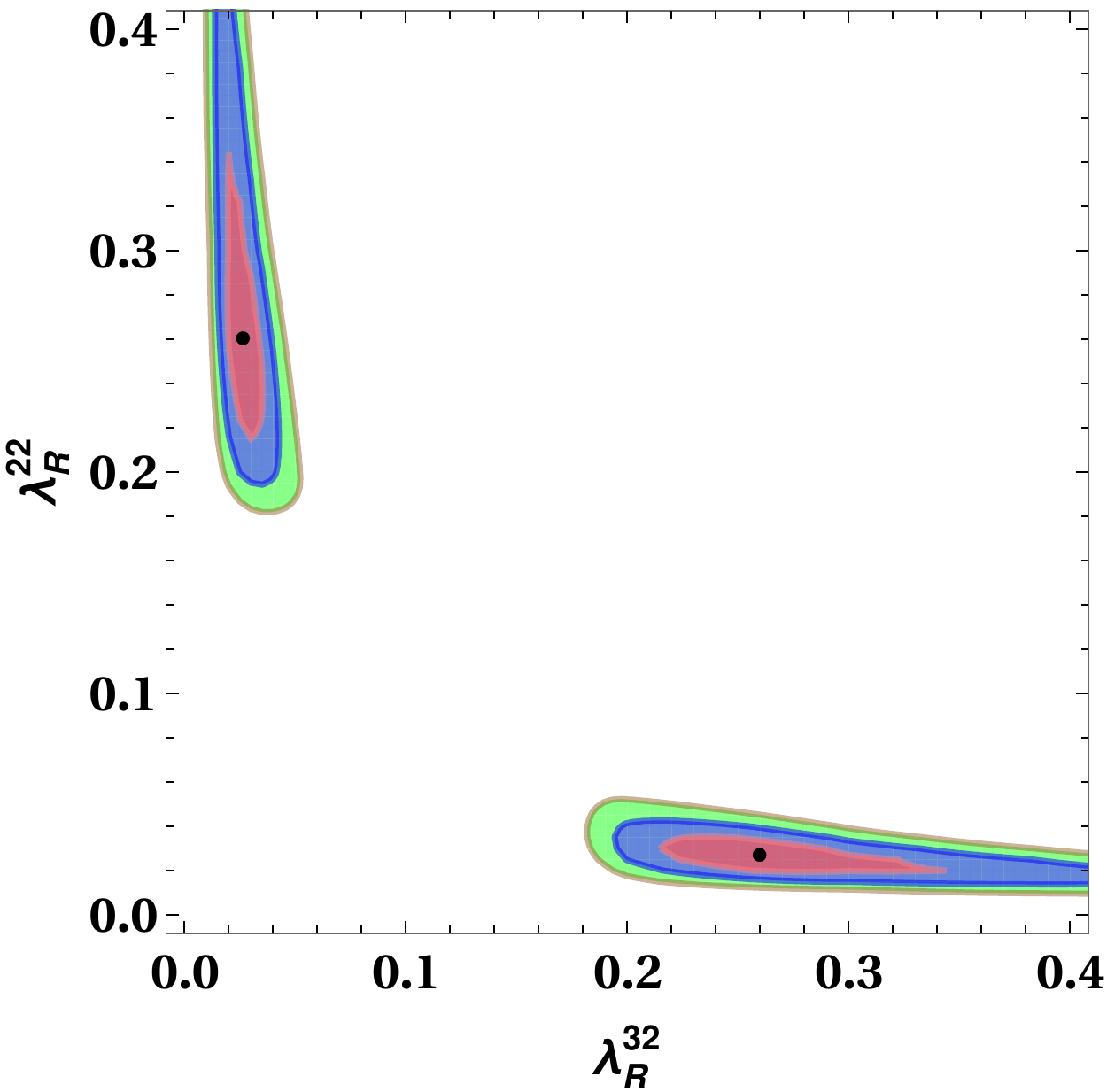}}
\caption{ Constraints on new LQ  couplings from observables mediated by $b \to c \tau \bar \nu$ (left panel) and  $b \to s \mu^+  \mu^-$ (right panel). Different colors correspond to $1\sigma$, $2\sigma$, and $3\sigma$ contours and the black dots represent  the best-fit value.} \label{Fig:S-I}
\end{figure*}

Here, we consider the NP contributions to both neutral current $b \to s \ell \ell$ as well as charged current $b \to c \tau \bar \nu_\tau$ processes, and constrain the NP couplings by confronting the SM results  with their corresponding observed  data.  In doing so, we perform the  $\chi^2$ analysis, wherein we use the following  expression  for our analysis
\begin{eqnarray} \label{Eq:chi}
\chi^2(C_i^{\rm NP}) \ = \ \sum_i \frac{\left[\mathcal{O}_i^{\rm th}(C_i^{\rm NP})-\mathcal{O}_i^{\rm exp}\right]^2}{(\Delta \mathcal{O}_i^{\rm exp})^2+(\Delta \mathcal{O}_i^{\rm th})^2}\;.
\end{eqnarray}
Here,  $\mathcal{O}_i^{\rm th}(C_{i}^{\rm NP})$ are the theoretically predicted values  for different  observables used in our fit, which are dependent  on the new Wilson coefficients $C_{i}^{\rm NP}$ and $\Delta \mathcal{O}_i^{\rm th}$ represent the $1\sigma$ uncertainties from theory inputs.   $\mathcal{O}_i^{\rm exp}$ and $\Delta \mathcal{O}_i^{\rm exp}$  illustrate the corresponding experimental  central values and their $1\sigma$ uncertainties.   In this analysis, we use  a represenative value of the LQ mass as $m_{\rm LQ}=1.2$ TeV, which is congruous with the  constraint obtained  from   LHC experiment~\cite{Sirunyan:2018kzh}. We further take into account the following two  scenarios   to obtain the best-fit  values of the LQ couplings. 
\begin{itemize}
\item \textbf{C-I} :  In this case, we include the observables associated with the charged current transitions of leptonic/semileptonic $B$ meson decays,  involving only third generation leptons, i.e.,  the processes mediated through $b \to c \tau \bar \nu_\tau$ transitions
\item \textbf{C-II} : Here, we incorporate the measurements on  leptonic/semileptonic $B$ decay modes involving only second generation leptons,  i.e., $b \to s \mu^+ \mu^-$ mediated processes.
\end{itemize}
In left panel of Fig. \ref{Fig:S-I}\,, we display the constraints on  the leptoquark couplings, which are obtained  by using the observables associated with $b \to c \tau \bar \nu$ transitions and  the plot on the right panel demonstrates the constraints obtained from $b \to s \mu^- \mu^+$ observables. Different colors in these plots  symbolize the $1\sigma$, $2\sigma$, and $3\sigma$ contours and the black dots represent the best-fit values.    The  best-fit values for the LQ couplings obtained for these two cases are presented in Table \ref{Tab:bestfit} along with their corresponding  pull values, defined as:  pull$=\sqrt{\chi_{\rm SM}^2-\chi^2_{\rm best-fit}}$.  
%
\begin{table}[t!]
\begin{center}
\small{\begin{tabular}{|c|c|c|c|}
\hline
~Scenarios~&~~Couplings~ &~Best-fit Values~&~Pull~\\
\hline
\hline

~C-I~&~~$(\lambda_L^{23},~{\lambda_R^{33}})$~ &~ $(-0.143,0.147)$ ~~&~$3.5$~\\ 
~~&~~~ &~$(0.147, -0.143)$ ~~&~~\\ 
\hline
~C-II~&~~$(\lambda_R^{32},~{\lambda_R^{22}})$~ &~~$(0.0265,0.260)$ ~~&~$5.4$~\\ 
~~&~~~ &~$(0.260,0.0265)$ ~~&~~\\ 
\hline
\end{tabular}}
\end{center}
\caption{. Best-fit values of new LQ couplings, and pull values  for all cases (C-I, C-II). } \label{Tab:bestfit}
\end{table}

 \subsection{Implications on lepton flavor violating $B$ and $\tau$ decays} \label{sec:LFV}
\begin{figure}[t!]
	\centering
	\hspace*{-0.5cm}
	\includegraphics[width=0.7\textwidth]{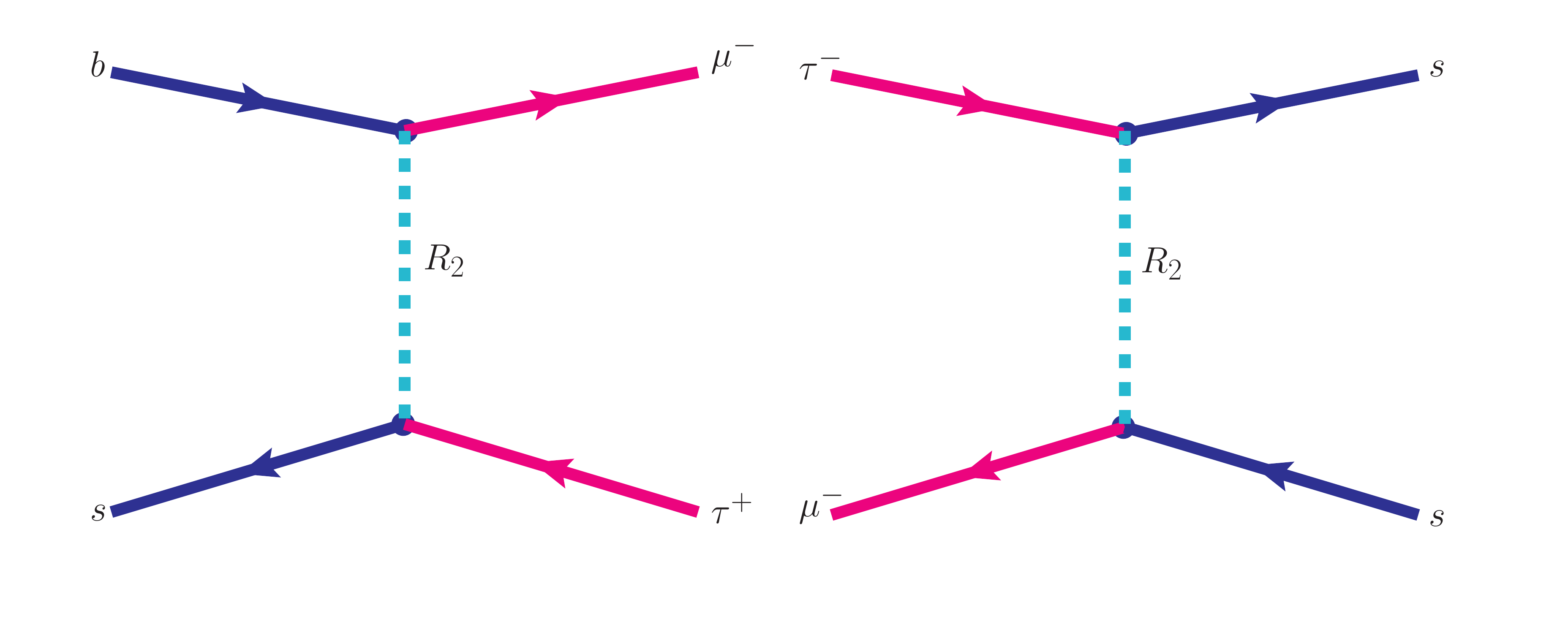}
	\caption{Feynman diagrams for the LFV processes  $b \to s \tau^+ \mu^- $  (left panel), and $\tau \to \mu \phi ~(\eta^{(\prime)})$ (right panel) mediated through  the exchange of scalar LQ.}
		\label{fig:Fyn-LFV}
\end{figure}

In this section, we will discuss some of the lepton flavor violating (LFV) decay modes
$B_{(s)}$ and $\Upsilon$ mesons as well as  $\tau$ lepton, due to the impact of  the scalar leptoquark, $R_2({\bf 3,2},7/6)$.   The rare leptonic/semileptonic LFV decays  of $B$ mesons involving the quark-level transitions $b \to s \ell_i^+ \ell_j^-$,  occur at tree level via the exchange of the SLQ. For illustration, in the left panel of  Fig. \ref{fig:Fyn-LFV}, we show  the Feynman diagram for $b \to s \tau \mu$ LFV process as a typical example. The  effective Hamiltonian for $b \to s \ell_i^+\ell_j^-$ process  due to the effect of  scalar LQ can be given   as \cite{Sahoo:2015wya, Sahoo:2015pzk}
\bea
\mathcal{H}_{\rm eff}\left( b \to s \ell_i^+ \ell_j^- \right)  &\ = \ & \Big[ G_V \left( \bar{s} \gamma^\mu P_L b \right) \left(\bar{\ell}_i \gamma_\mu \ell_j\right) + G_A \left( \bar{s} \gamma^\mu P_L b \right) \left(\bar{\ell}_i \gamma_\mu \gamma_5 \ell_j\right) \Big]\;,
\eea 
where the vector and axial vector couplings $G_{V,A}$ are expressed as
\bea
G_V=G_A =\frac{\lambda_R^{j3} (\lambda_R^{i2 })^*}{8 m_{\rm LQ}^2}\;.
\eea
This effective Hamiltonian leads to the following decay processes:
\begin{enumerate}

\item {$\boldsymbol{B_s \to \ell_i^+ \ell_j^-}$} : The branching ratio of the LFV decay process $B_s\to \ell_i^+ \ell_j^-$, in the presence of scalar LQ  is given as ~\cite{Becirevic:2016zri}
\begin{align}\label{Eq:LFV-Bsll}
& {\rm BR}(\overline{ B}_s\to  \ell_i^- \ell_j^+) \ = \ \tau_{B_s} \frac{1}{8 \pi M_{B_s}^3}|f_{B_s}G_V|^2 \lambda^{1/2}(M_{B_s}^2,m_i^2,m_j^2)\nonumber\\
&\times\Bigg [(m_j-m_i)^2 \Big (M_{B_s}^2-(m_i+m_j)^2\Big )  
+ (m_j+m_i)^2 \Big(M_{B_s}^2-(m_i-m_j)^2 \Big )  \Bigg ],
\end{align}
where  $f_{B_s}$ represents the decay constant of  $B_s$ and 
\bea
\lambda(a,b,c) \ = \ a^2 + b^2 +c^2 -2\left( a b + b c + a c \right) \;,
\label{eq:triangle}
\eea
is the triangle  function. 

\item {$\boldsymbol{\overline{ B} \to \overline{ K} \ell_i^+\ell_j^-}$} : The differential branching fraction of $B \to K \ell_i^+ \ell_j^-$ process is given as~\cite{Sahoo:2015pzk}
\bea \label{Eq:LFV-BKll}
\frac{{\rm d}{\rm BR}}{{\rm d}q^2}(\overline{B}\to \overline{K}\ell_i^+ \ell_j^-)  =  2\tau_B  \Big (a(q^2) + \frac{1}{3} c(q^2) \Big )\,,
\eea
where the coefficients $a(q^2)$ and $c(q^2)$ are  expressed as
\bea
a (q^2) & =& \Gamma_0 \frac{\sqrt{\lambda_1 \lambda_2}}{q^2} f_+^2 \Bigg[ \left(|G_V|^2 + |G_A|^2 \right) \frac{\lambda_1}{4}
 + |G_S|^2 \left (q^2 - (m_i + m_j)^2\right )\nn\\ 
& + & |G_P|^2 \left (q^2 - (m_i - m_j)^2 \right )  + |G_A|^2 M_B^2 (m_i + m_j)^2 + |G_V|^2 M_B^2 (m_i - m_j)^2\nn  \\
& + & \left(M_B^2 - M_K^2 + q^2 \right) \Big( (m_i + m_j)  Re (G_P G_A^*) + (m_j - m_i)  Re (G_S G_V^*)\Big) \Bigg]\;,
\eea
\begin{equation}
c (q^2) = - \Gamma_0  f_+^2 \frac{(\lambda_1 \lambda_2)^{3/2}}{4 q^6} \left(|G_A|^2 + |G_V|^2 \right)\;, \hspace{6cm}
\end{equation}
with
\bea
\Gamma_0  =  \frac{1}{2^8 \pi^3 M_B^3}, \hspace{1cm} \lambda_1 = \lambda (M_B^2, M_K^2, q^2),  \hspace{1cm} \lambda_2 = \lambda (q^2,m_i^2, m_j^2)\;,
\eea
and 
\bea
G_S &= & \frac{1}{2} G_{V} (m_j - m_i) \Bigg[\frac{M_B^2 - M_K^2}{q^2}  \left( \frac{f_0 (q^2)}{f_+ (q^2)} - 1 \right) -1 \Bigg]\;,\nn\\
G_P & = & \frac{1}{2} G_{A} (m_i + m_j)\left[\frac{M_B^2 - M_K^2}{q^2} \left( \frac{f_0 (q^2)}{f_+ (q^2)} - 1 \right)-1 \right]\;.
\eea
$f_{0,+}$ are the form factors describing $B \to K$ transitions.
\item {$\boldsymbol{\overline{B }\to \overline{K}^* \ell_i^+ \ell_j^-}$ \textbf{and} $\boldsymbol{\overline{B}_s \to \phi \ell_i^+ \ell_j^-}$} : The differential branching fraction of $\overline{B}\to  \overline{K}^* \ell_i^+ \ell_j^-$  process is given as~\cite{Sahoo:2015pzk}
\bea
\frac{\rm d BR}{dq^2} & = & \tau_B \Gamma_V \times \Bigg[ A(q^2)^2 \Bigg\{ \frac{2}{3} \lambda_{K^*} \left(1-\left( \frac{m_i^2}{q^2} \right)^2\right) 
+  8M_{K^*}^2 (q^2 - m_i^2) \nn \\ && -\frac{2}{9} \left(1-\frac{m_i^2}{q^2}  \right)^2 \left( \left(M_B^2-M_{K^*}^2-q^2\right)^2 
+ 8q^2 M_{K^*}^2 \right) \Bigg\} \nn \\ && + B(q^2)^2 \Bigg\{\frac{\lambda_{K^*}}{6}\left(M_B^2-M_{K^*}^2-q^2\right)^2 \left(1-
\left( \frac{m_i^2}{q^2} \right)^2\right) -\frac{\lambda_{K^*}^2}{18}\left(1-\frac{m_i^2}{q^2}  \right)^2 \nn \\ &&
 -\frac{2}{3}\lambda_{K^*}M_{K^*}^2 (q^2 - m_i^2) \Bigg\}  +C(q^2)^2 \Bigg\{ \frac{2}{3} \lambda_{K^*}m_i^2(q^2 - m_i^2) \Bigg\}\nn \\
 && -D(q^2)^2 \Bigg\{ \frac{4}{9}\lambda_{K^*}M_{K^*}^2(q^2 - m_i^2)\left(4-\frac{m_i^2}{q^2}  \right) \Bigg\} \nn \\ && 
-Re\Big (A(q^2)B(q^2)^*\Big )\Bigg\{ \frac{2}{3}\lambda_{K^*}\left(M_B^2-M_{K^*}^2-q^2\right)\left(1-\left( \frac{m_i^2}{q^2} \right)^2
-\frac{1}{3}\left(1-\frac{m_i^2}{q^2}  \right)^2\right) \Bigg\}  \nn \\ 
&& -Re\Big(A(q^2)C(q^2)^*\Big) \Bigg\{\frac{4}{3}\lambda_{K^*}m_i^2\left(1-\frac{m_i^2}{q^2} \right)\Bigg\}  \nn \\
 && +Re\Big(B(q^2)C(q^2)^*\Big)\Bigg\{ \frac{2}{3}\lambda_{K^*}m_i^2\left(M_B^2-M_{K^*}^2-q^2\right)\left(1-\frac{m_i^2}{q^2} \right)\Bigg\} \Bigg],
\eea
where
\bea
\Gamma_V = \frac{3\sqrt{\lambda_{K^*}}}{2^{11}M_{K^*}^2\left(\pi M_B \beta \right)^3} |G_{V}|^2\;, \hspace{0.5cm} 
\lambda_{K^*} = \lambda \left(M_B^2, M_{K^*}^2, q^2\right)\;, \hspace{0.5cm} 
 \beta ={1\over M_{K^*}^2}\lambda^{1/2} \left(M_B^2, M_{K^*}^2, q^2\right),\nn\\
\eea
and the functions $A(q^2),~B(q^2),~C(q^2)$ and $D(q^2)$ are related to the various form factors of $B \to K^*$ transitions as
\bea
A(q^2)&=&\left(M_B+M_{K^*}\right)A_1(q^2)\;, \hspace{2.8cm} B(q^2)=\frac{2A_2(q^2)}{\left(M_B+M_{K^*}\right)}\;, \nn \\  
C(q^2)&=&\frac{A_2(q^2)}{\left(M_B+M_{K^*}\right)}+\frac{2M_{K^*}}{q^2}\left(A_3(q^2)-A_0(q^2)\right)\;, 
\hspace{0.25 cm} D(q^2)=\frac{2V(q^2)}{\left(M_B+M_{K^*}\right)}\;.\hspace*{0.5 true cm}
\eea
The same expression can be used for $B_s \to \phi \ell_i \ell_j$ processes by appropriately replacing the  particle masses and the lifetime of $B_s$ meson. For numerical estimation, we use  the particle masses and  $B$ meson  lifetimes as well as other input parameters  from PDG~\cite{Zyla:2020zbs}\,. Using $f_{B_s}=(225.6\pm 1.1\pm 5.4)$ MeV \cite{Charles:2015gya} and    best-fit values of the  new couplings from Table \ref{Tab:bestfit}, we present  our predicted results on  various branching ratios of LFV decays of $B$ mesons in Table~\ref{Tab:LFV}\,. It can be noticed from the table that  the branching fractions of various LFV $B$ decays are quite significant in the presence of $R_2$ scalar leptoquark and  are within the reach of Belle-II or LHCb experiments. However,  for most of these decays,  the experimental limits   are  not yet available. The  LFV channels which have been searched for  are $B^+ \to K^+ \mu^-\tau^+ (\mu^+\tau^-) $~\cite{Lees:2012zz} and $B_s\to \tau^\pm \mu^\mp$~\cite{Aaij:2019okb} for which we find our predicted branching fraction values are well below  the present 90\% CL upper limits. Our obtained result on  ${\rm BR}(B_s\to \tau^\pm \mu^\mp)$ is 
\bea
{\rm BR}(B_s\to \tau^\pm \mu^\mp) \ =  \ {\rm BR}(B_s\to \tau^+ \mu^-)+{\rm BR}(B_s\to \tau^- \mu^+) = 
1.3\times 10^{-9} \,,
\eea 
which is well below the current  experimental limit at 90\% C.L.~\cite{Aaij:2019okb}
$
{\rm BR}(B_s\to \tau^\pm \mu^\mp)^{\rm exp} \ < \ 3.4\times 10^{-5} \,.
$
Our predicted branching ratios for the LFV processes   $B_{(s)} \to (K, K^*, \phi) \mu^-\tau^+ (\mu^+\tau^-)$  are quite reasonable and are within the reach of  Belle-II~\cite{Kou:2018nap} as well as the upcoming  LHCb upgrade~\cite{Bediaga:2018lhg}. In Fig. \ref{Fig:LFV}\,, we display  the differential branching fractions of the decay modes  $B^+ \to K^+ \mu^- \tau^+$  (top-left panel), $B^+ \to K^{*+} \mu^- \tau^+$ (top-right panel) and $B_s \to \phi \mu^- \tau^+$ (bottom panel) with respect to $q^2$. 
\begin{table*}[t!]
\begin{center}
\begin{tabular}{|c|c|c|}
\hline
Decay  modes &Predicted values & Experimental Limit  \\ 

\hline
\hline
$B_s \to \mu^- \tau^+$~&~$3.03\times 10^{-9}$~~&~$<3.4\times 10^{-5}$~(90\% CL) \cite{Aaij:2019okb}\\

$B^+ \to K^+ \mu^- \tau^+$~&~$1.5 \times 10^{-8}$~&~~$<2.8\times 10^{-5}$~(90\% CL) \cite{Lees:2012zz}\\

$\overline B^0 \to \overline K^0 \mu^- \tau^+$~&~$1.4 \times 10^{-8}$~&~$\cdots$\\

$B^+ \to K^{* +} \mu^- \tau^+$~&~$2.91\times 10^{-8}$~&~$\cdots$\\

$\overline B^0 \to \overline K^{* 0} \mu^- \tau^+$~&~$2.7\times 10^{-8}$~&~$\cdots$\\

$B_s \to \phi \mu^- \tau^+$~&~$3.5\times 10^{-8}$~&~$\cdots$\\

\hline
$B_s \to \mu^+ \tau^-$~&~$3.2\times 10^{-9}$~&~$<3.4\times 10^{-5}$~(90\% CL) \cite{Aaij:2019okb}\\
$B^+ \to K^+ \mu^+ \tau^-$~&~$1.6\times 10^{-8}$~&~$<4.5\times 10^{-5}$~(90\% CL) \cite{Lees:2012zz}\\
$\overline B^0 \to \overline K^0 \mu^+ \tau^-$~&~$1.5\times 10^{-8}$~&~$\cdots$\\
$B^+ \to K^{* +} \mu^+ \tau^-$~&~$3.1\times 10^{-8}$~&~$\cdots$\\
$\overline B^0 \to \overline K^{* 0} \mu^+ \tau^-$~&~$2.8\times 10^{-8}$~&~$\cdots$\\
$B_s \to \phi \mu^+ \tau^-$~&~$3.7\times 10^{-8}$~&~$\cdots$\\
\hline
$\Upsilon(1S) \to  \mu^\mp \tau^\pm $~&~$2.12\times 10^{-12}$~&$6.0\times 10^{-6}~(95\% ~{\rm CL})$ \cite{Zyla:2020zbs}\\
$\Upsilon(2S) \to  \mu^\mp \tau^\pm $~&~$2.16\times 10^{-12}$~&$3.3\times 10^{-6}~(90\% ~{\rm CL})$ \cite{Zyla:2020zbs}\\
$\Upsilon(3S) \to  \mu^\mp \tau^\pm$~&~$2.82\times 10^{-12}$ &$3.1 \times 10^{-6}~(90\% ~{\rm CL})$ \cite{Zyla:2020zbs}\\
\hline
$\tau^- \to\mu^- \phi$ ~&~$4.4\times 10^{-10}$  ~&~$<8.4\times 10^{-8}$~(90\% CL) \cite{Miyazaki:2011xe}\\
$\tau^- \to\mu^- \eta$ ~&~$2.18\times 10^{-10}$ & ~$<6.5\times 10^{-8}$~(90\% CL) \cite{Zyla:2020zbs}\\
$\tau^- \to\mu^- \eta^\prime$ ~&~$5.49\times 10^{-10}$  ~&~$<1.3\times 10^{-7}$~(90\% CL) \cite{Zyla:2020zbs}\\

\hline
\end{tabular}
\end{center}
\caption{Predicted values of the  branching ratios of lepton flavor violating decay channels of $B$  meson and $\tau$ lepton in the present model.} 
 \label{Tab:LFV}
\end{table*}

\begin{figure*}[t!]
\centering
\subfigure[~$B^+ \to K^+ \mu^-\tau^+$]{\includegraphics[width=0.4\textwidth]{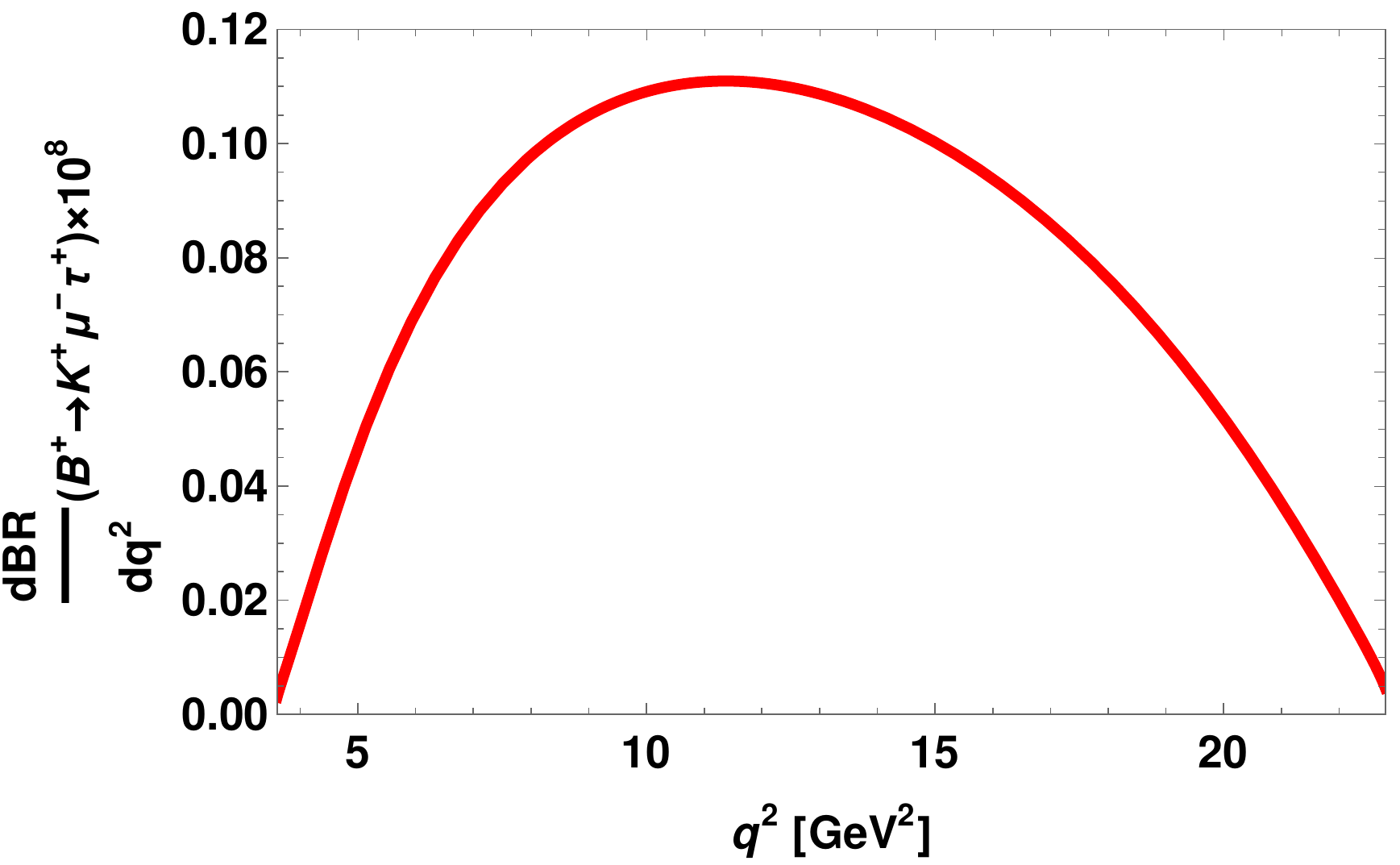}}
\quad
\subfigure[~$B^+ \to K^{*+} \mu^-\tau^+$]{\includegraphics[width=0.4\textwidth]{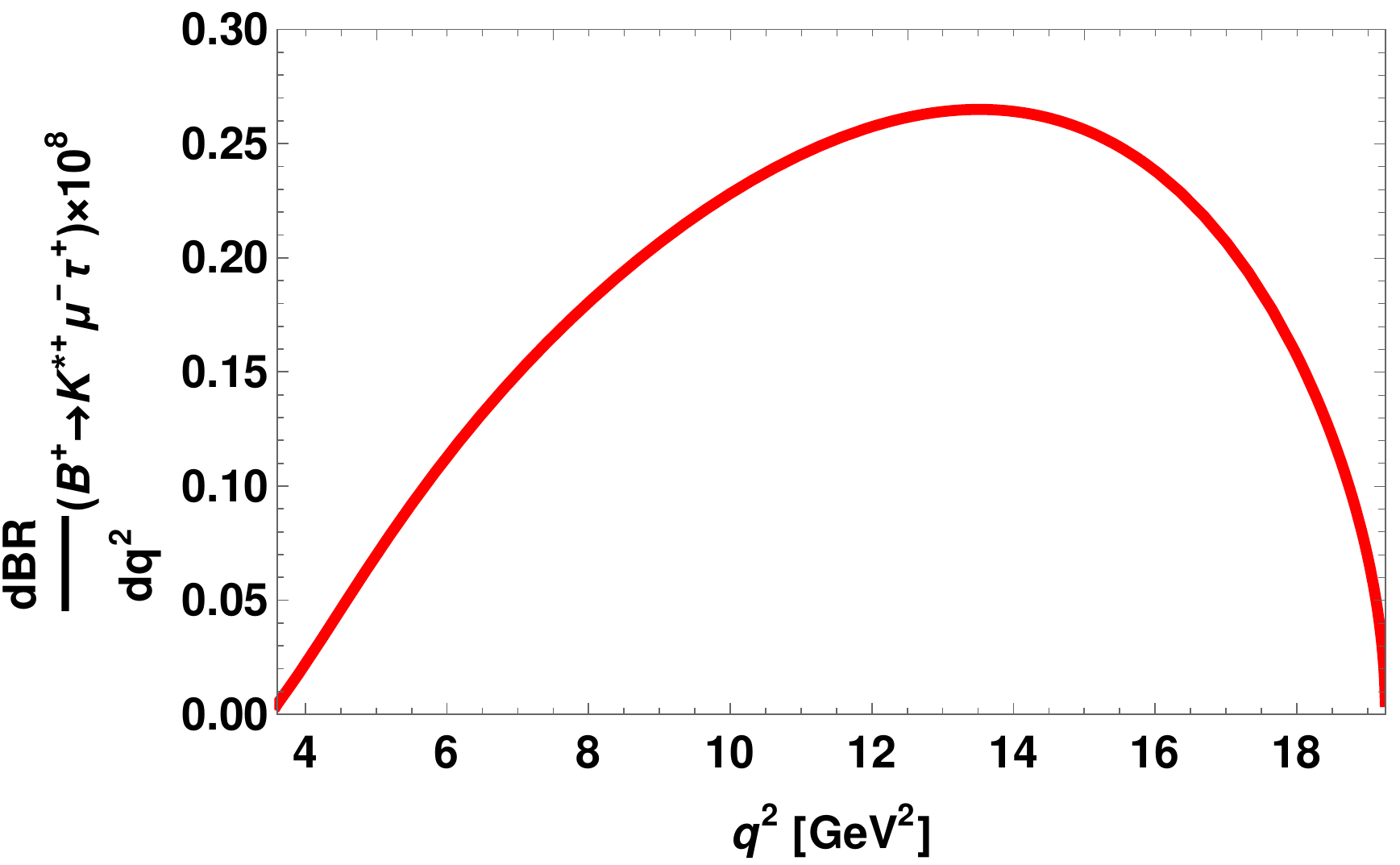}}
\quad
\subfigure[~$B_s \to \phi \mu^-\tau^+$]
{\includegraphics[width=0.4\textwidth]{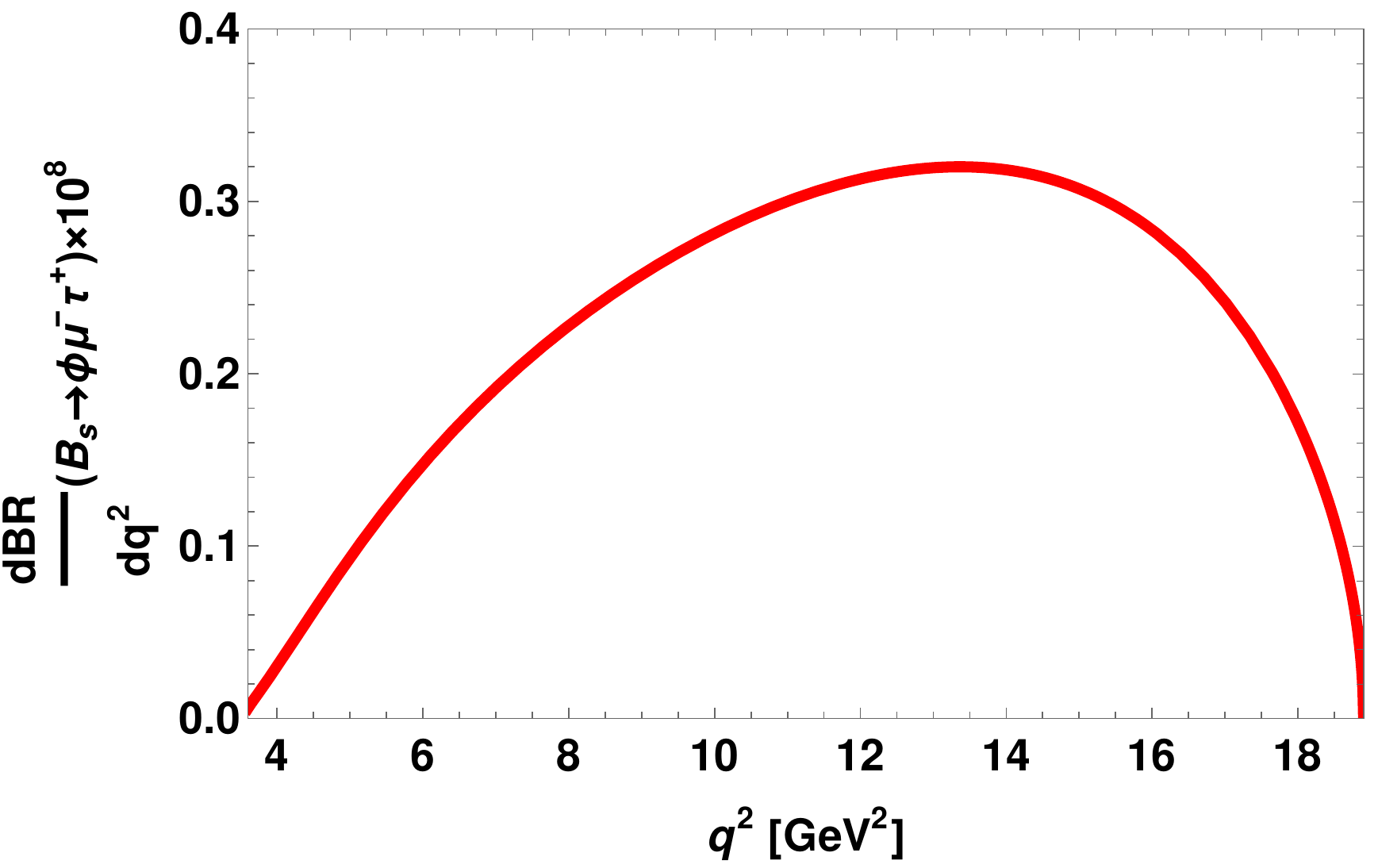}}
\caption{Behaviour of  the differential branching fractions of the LFV processes (a) $B^+ \to K^+ \mu^- \tau^+$,  (b)$B^+ \to K^{*+}\mu^- \tau^+$  and (c) $B_s \to \phi \mu^- \tau^+$ with respect to $q^2$ due to the effect of $R_2^{3/2}$ leptoquark. } \label{Fig:LFV}
\end{figure*}

\item {${\boldsymbol{\Upsilon(nS) \to \mu \tau}}$} : The  LFV process  $\Upsilon(nS) \to \mu \tau$ can occur at tree level in the LQ model and the corresponding Feynman diagram can be obtained from that of  $b \to s \mu \tau$ process (left panel of Fig. \ref{fig:Fyn-LFV}) by replacing $s\to b$, and  the branching ratio for this process  is given as  \cite{Bhattacharya:2016mcc}
\bea
\label{Eq:upsilontaumu}
{\rm BR}(\Upsilon(nS) \to  \mu^- \tau^+) \ & = & \ \frac{f_{\Upsilon(nS)}^2 m_{\Upsilon(nS)}^3}{48 \pi \Gamma_{\Upsilon(nS)} } \left(2-\frac{m_\tau^2}{m_{\Upsilon(nS)}^2}-\frac{m_\tau^4}{m_{\Upsilon(nS)}^4}\right)\left(1-\frac{m_\tau^2}{m_{\Upsilon(nS)}^2}\right)\left|\frac{\lambda_{R}^{32}\lambda_{R}^{33*}}{8 m_{{\rm LQ}}^2}\right|^2\,.\nn\\
\eea
The branching ratio  for the  process $\Upsilon(nS) \to \mu^+ \tau^-$  can be obtained from ${\rm BR}(\Upsilon(nS) \to \mu^- \tau^+)$ by appropriately replacing the LQ couplings, i.e.,  $\lambda_{R}^{32}\lambda_{R}^{33*}\to \lambda_{R}^{33}\lambda_{R}^{32*}$. Hence, the branching ratio for $\Upsilon(nS) \to  \mu^\mp \tau^\pm$ process is given as
\bea
{\rm BR}(\Upsilon(nS) \to  \mu^\mp \tau^\pm)={\rm BR}(\Upsilon(nS) \to  \mu^- \tau^+)+ {\rm BR}(\Upsilon(nS) \to  \mu^+ \tau^-)\,.
\eea
For numerical estimation,  all the particle masses and  widths of  $\Upsilon(nS),~n=1,2,3$ are taken from PDG \cite{Zyla:2020zbs}\,. The values of $\Upsilon(nS)$ decay constants used are as follows: $f_{\Upsilon(1S)}=(700\pm 16)$ MeV, $f_{\Upsilon(2S)}=(496\pm 21)$ MeV and $f_{\Upsilon(3S)}=(430\pm 21)$ MeV \cite{Bhattacharya:2016mcc}\,.   With these input parameters the predicted  branching ratios of $\Upsilon(nS) \to  \mu^\pm \tau^\mp$ are provided in Table \ref{Tab:LFV}\,,
which are far below the current experimental upper limits~\cite{Zyla:2020zbs}.

\item {${\boldsymbol{\tau \to \mu \phi}}$} : The Feynman diagram for the LFV decay process $\tau \to \mu \phi$  is presented in the right panel of Fig. \ref{fig:Fyn-LFV} and its   branching ratio is expressed as \cite{Becirevic:2016oho}
\bea
\label{Eq:taumuphi}
{\rm BR}(\tau\to  \mu \phi) \  =  \ \frac{\tau_\tau f_\phi^2 m_\phi^4}{256 \pi m_\tau^3}\left|\frac{\lambda_{R}^{23}\lambda_{R}^{22*}}{m_{{\rm LQ}}^2}
\right|^2  
\times   \lambda^{1/2}(m_\phi^2,m_\tau^2,m_\mu^2) \left[-1+\frac{(m_\mu^2+m_\tau^2)}{2
    m_\phi^2}+\frac{(m_\mu^2-m_\tau^2)^2}{2
    m_\phi^4}\right],\nonumber\\
\eea
where $f_\phi$  is the $\phi$ meson decay constant.  
 Using $f_\phi=(238\pm 3)$ MeV from Ref.~\cite{Chakraborty:2017hry}, and the other input parameters   from PDG~\cite{Zyla:2020zbs},  along with the best-fit values of required new parameters from Table \ref{Tab:bestfit}\,,  the predicted branching fraction of $\tau \to \mu \phi$ is shown in Table~\ref{Tab:LFV}\,.  We find that the branching ratio is  substantially enhanced and is within the reach of Belle-II experiment.

\item {${\boldsymbol{\tau \to \mu \eta^{(\prime)}}}$} : The branching ratio for $\tau \to \mu \eta^{(\prime)}$  process is given as 
\bea
\label{Eq:taumueta}
{\rm BR}(\tau\to  \mu \eta^{(\prime)}) \ & = & \ \frac{\tau_\tau f_{\eta^{(\prime)}}^2 m_{\tau}^3}{512 \pi }\left|\frac{\lambda_{R}^{23}\lambda_{R}^{22*}}{m_{{\rm LQ}}^2}\right|^2   \left(1-\frac{m_{\eta^{(\prime)}}^2}{m_\tau^2} \right)^2\,.
\eea
Using $f_\eta\simeq -157.63$ MeV, \cite{Bhattacharya:2016mcc}, $f_{\eta^\prime}\simeq 31.76$ MeV  \cite{Bhattacharya:2016mcc}, along with other input parameters  from \cite{Zyla:2020zbs} and the best-fit values of LQ couplings from Table \ref{Tab:bestfit}\,, our predicted  values of branching ratios of $\tau \to \mu \eta^{(\prime)}$ processes are presented in Table \ref{Tab:LFV}\,, which are found to be substantially lower than the current experimental upper limits. 


\end{enumerate}

\section{Dark Matter}
We consider fermion triplet $\Sigma(1,3,0)$ coming from the fermion representation $45_F$ of $SO(10)$. The stability of fermion triplet dark matter is ensured from the matter parity under which $16_F$ is odd while $45_F$ is even. SM Higgs is contained in $10_S$ and the scalar leptoquark $R_2$ is contained in $126_S$, are both even under matter parity  \cite{Hambye:2010zb}. The generic Yukawa term $y \Sigma \bar{\ell_L} \phi$ mediating neutrino masses by type-III seesaw is not allowed, which can be understood as follows. The SM lepton doublet is contained in $16_F$, scalar doublet resides in $10_S$, while the fermion triplet DM exists in $45_F$ and the Lagrangian term in $SO(10)$ bilinear $16_F 10_H 45_F$ is actually forbidden because of the matter parity. Hence the fermion triplet mass comes from the invariant bilinear $M_{\Sigma} 45_F 45_F$ and the relic density of DM is solely controlled by the gauge interactions. The low energy invariant interaction term for fermion triplet DM is given by
\begin{eqnarray}
\mathcal{L}_{\rm \Sigma} &= \frac{i}{2}{\rm Tr}[\overline{\Sigma_R}\slashed{D}\Sigma_R]+\frac{i}{2}{\rm Tr}[\overline{\Sigma^c_R}\slashed{D}\Sigma^c_R] - \left(\frac{1}{2}{\rm Tr}[\overline{\Sigma^c_R}M_\Sigma\Sigma_R] + {\rm h.c.}\right),
\end{eqnarray}
where, $\Sigma^c_R = C \overline{\Sigma_R}^T$ is the CP conjugate of $\Sigma_R$ with $C$ being the operator for charge conjugation and $D_\mu$ is the covariant derivative for $\Sigma_R$, given by
\begin{equation}
D_\mu = \d_\mu \Sigma_R + ig \left[\sum_{a=1}^3 \frac{\sigma^a}{2}W^a_\mu, \Sigma_R\right]. 
\end{equation}
Defining the four component Dirac spinor as $\psi^- = \Sigma^-_R + \Sigma^{+c}_R$ and Majorana fermion as $\psi^0 = \Sigma^0_R + \Sigma^{0c}_R$, we write the Lagrangian for fermion triplet as \cite{Biswas:2018ybc}
\begin{eqnarray}
\mathcal{L}_{\rm triplet} &&= \overline{\psi^-}i\slashed{\d}\psi^- + \frac{1}{2}\overline{\psi^0}i\slashed{\d}\psi^0 - M_{\psi^-}\overline{\psi^-}\psi^- - \frac{M_{\psi^0}}{2}\overline{\psi^0}\psi^0 \nonumber \\
&&  + g\left(\cos \theta_w \overline{\psi^-}\gamma_\mu\psi^-Z^\mu + \sin \theta_w \overline{\psi^-}\gamma_\mu\psi^-A^\mu \right) \nonumber\\ &&-g\left(\overline{\psi^-}\gamma^\mu\psi^0 W_\mu^-+ {\rm h.c.}\right).
\label{FT_gauge}
\end{eqnarray}

\subsection{Relic abundance}
\begin{figure}[h]
\centering
\includegraphics[scale=0.54]{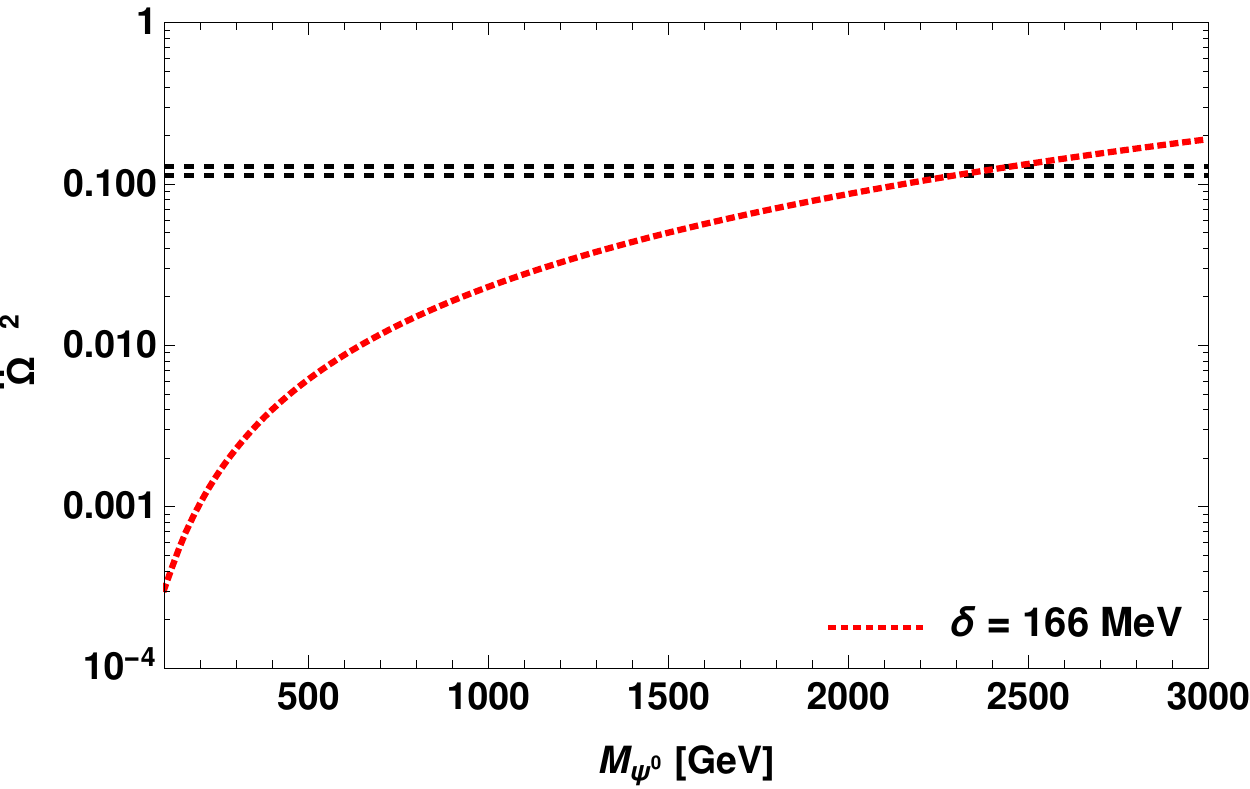}
 \caption{Relic abundance as a function of DM mass with contributions from annihilation's and co-annihilation's of $\psi^0$ and $\psi_\pm$. Black horizontal dashed lines correspond to $3 \sigma$ region of Planck satellite data. }
 \label{relic}
\end{figure}

The neutral component of fermion triplet ($\psi^0$) is Majorana type and the charged component ($\psi^{\pm}$) is Dirac in nature. At tree-level, both the charged and neutral components remain degenerate in mass. However, one-loop electroweak radiative corrections provide a mass splitting of $\delta = 166$ MeV \cite{Cirelli:2005uq,Ma:2008cu}, where $\delta = M_{\psi^\pm} - M_{\psi^0}$. Thus the Majorana fermion $\psi^0$ is the lightest thermal dark matter candidate in the present model and its relic density is governed by the gauge interactions  \ref{FT_gauge}.   
We have used the packages LanHEP \cite{Semenov:1996es} and micrOMEGAs \cite{Pukhov:1999gg, Belanger:2006is, Belanger:2008sj} to extract compute dark matter relic density. With the mentioned mass splitting, co-annihilation's  also contribute to dictate relic density in addition to annihilation's. The processes include $\psi^0 \overline{\psi^0} \to W^+ W^-$ (via t-channel $\psi^-$ exchange), $\psi^\pm \psi^\pm \to  W^\pm W^\pm$ (via t-channel $\psi^0$ exchange), $\psi^+ \psi^- \to  f \overline{f}$ (via s-channel $A,Z$ exchange) with $f = u,d,s,c,t,b,e,\mu,\tau$ and $\psi^0 \psi^- \to  \overline{f^\prime},f^{\prime\prime}$ (via s-channel $W^-$ exchange) with $f^\prime = u,c,t,\nu_e,\nu_\mu, \nu_\tau$ and $f^{\prime\prime} = d,s,c,e,\mu,\tau$. Fig. \ref{relic} depicts the relic density as a function of DM mass, with contribution from the above mentioned channels. The abundance meets the Planck limit ($3\sigma$) \cite{Aghanim:2018eyx} in the mass region $2.34$ TeV to $2.4$ TeV \cite{Ma:2008cu,Biswas:2018ybc}. 
\subsection{Direct searches}
\begin{figure}[h]
\centering
\includegraphics[scale=0.54]{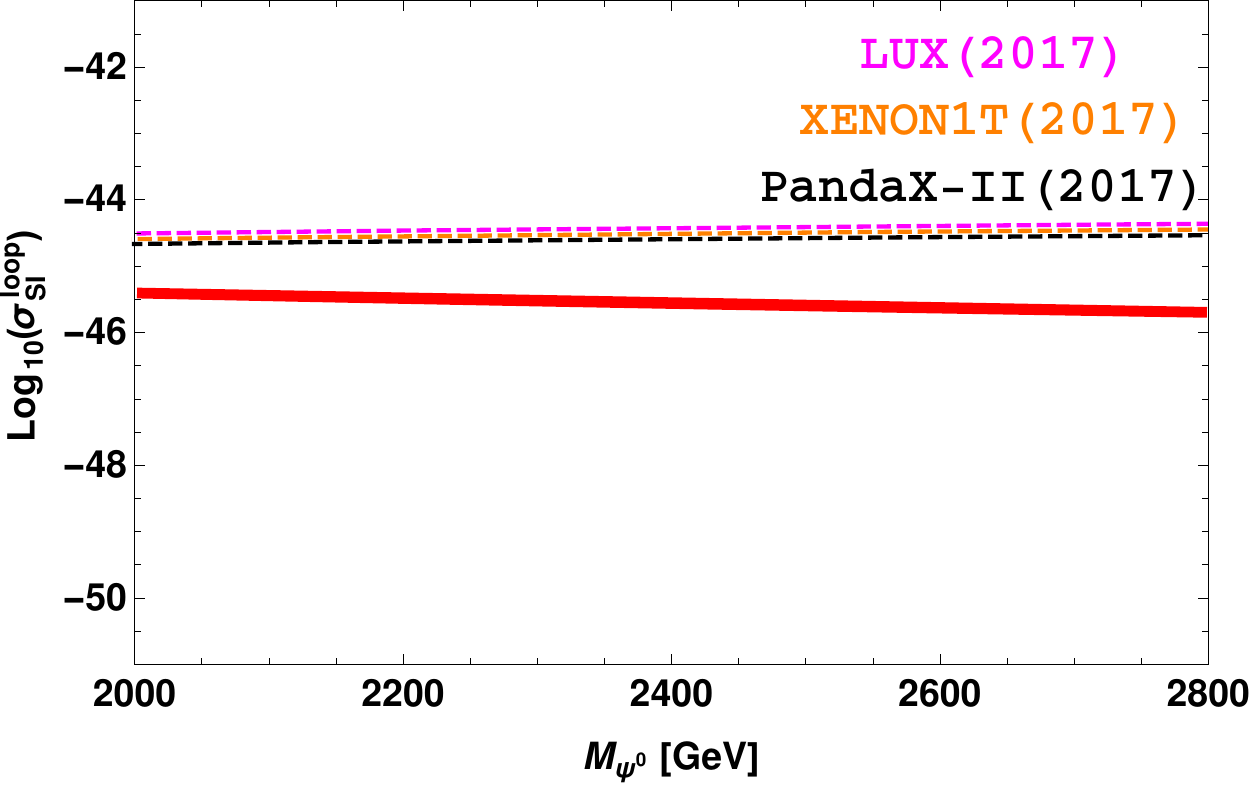}
\caption{One-loop SI contribution projected with the stringent upper limits of PandaX-II \cite{Cui:2017nnn}, XENON1T \cite{Aprile:2017iyp} and LUX \cite{Akerib:2016vxi}.}
\label{DD_loop}
\end{figure}

Moving on to the detection perspective, the neutral component $\psi^0$ can produce a nuclear recoil through Higgs penguin and box diagram  with W loop \cite{Cai:2015zza}. The effective interaction is given by
\begin{equation}
\mathcal{L} = \sum_i \xi^i_q \overline{\psi^0}\psi^0 \overline{q_i}q_i.
\end{equation}
Here, 
\begin{eqnarray}
\xi^i_q &=& -\alpha^2_2 \frac{m_{q_i}}{M_{\psi^0}} \bigg[-\frac{1-4n_W + 3n_W^2 - (4n_W-2){\rm log}~n_W}{m_h^2(1-n_W)^3} \nonumber\\
&& + \frac{2-3n_W+6n_W^2-5n_W^3+3n_W(1+n_W^2) {\rm log}~n_W}{6m_W^2(1-n_W)^4}\bigg],\nonumber\\
\end{eqnarray}
with $n_W = m_W^2/m^2_{\psi^0}$ and $\alpha_2 = \frac{g^2}{4\pi}$. Thus, the spin-independent (SI) cross section is given by
\begin{equation}
\sigma^{\rm loop}_{\rm SI} = \frac{4~ \mu_r^2}{\pi} m^2_{p} \left(\frac{\xi^i_q}{m_{q_i}}\right)^2 f_p^2,
\end{equation}
where, $m_p$ is proton mass, $\mu_r$ is the reduced mass of DM-nucleon system and $f_p \simeq 0.3$. Figure. \ref{DD_loop} projects the SI Cross section as a function of DM mass. We notice that the loop contribution is well below the upper limits levied by PandaX-II \cite{Cui:2017nnn}, XENON1T \cite{Aprile:2017iyp} and LUX \cite{Akerib:2016vxi}.

In case of indirect searches, DM can provide gamma ray signal via $W^\pm$ at loop level. However, Fermi-LAT with Sommerfeld enhancement rules out such suppressed cross section from a TeV scale DM \cite{Bern:1997ng,Choubey:2017yyn}.

\section{Conclusion}
We have considered an extension of standard model by a scalar leptoquark $R_2$ and a fermion triplet $\Sigma$ and embedded the framework in non-SUSY $SO(10)$ GUT. The introduction of $R_2$ and $\Sigma$ at few TeV scale assist the unification of gauge couplings of strong and electroweak forces while consistent with flavor anomalies $R_K$, $R_{K^{(*)}}$, $R_{D^{(*)}}$, $R_{J/\psi}$ and dark matter phenomenology.

The right-handed neutrino which is part of $16_F$ spinorial representation of $SO(10)$ can explain the non-zero neutrino masses. The dark matter comes from $45_F$ of $SO(10)$, while the scalar leptoquark is contained in the $126_S$. Since $16_F$ is odd while all other multiplets are even under matter parity $P_M$ and thus ensure the stability of fermion triplet dark matter.

The unification mass scale comes out to be $M_U= 10^{13.27}\,\,\mbox{GeV}$ which is well below the  limit set by the proton decay experiments. In order to satisfy experimental bound on proton decay, we adopt one loop GUT threshold corrections arising due to presence of super heavy scalars, fermions and gauge bosons by modifying the one-loop beta coefficients and revolution of gauge couplings at the GUT scale $M_U$. After including threshold corrections, the modified value of unification mass scale is found to be $10^{16.323}$, which resulted the proton life time as $\tau_p = 7.7 \times 10^{36}$ years.

The proposed model incorporates the scalar leptoquark $R_2(3,2,7/6)$, which plays a crucial role in explaining the recently observed flavor anomalies in  semileptonic $B$ decays. 
The intriguing feature of this leptoquark is that, it can induce  additional contributions to the CC $b \to c \tau \bar \nu_\tau$ as well as NC $b \to s \ell^+ \ell^-$ transitions at the tree level due to the exchange of LQ and hence, can successfully account for the observed discrepancies in the LFU violating observables. In this work, the leptoquark couplings  are constrained by using the LFU observables $R_{D^{(*)}}$, $R_{J/\psi}$, ${\rm BR} (B_c \to \tau \bar \nu_\tau)$ for $b \to c \tau \bar \nu_\tau$ transitions and $R_{K^{(*)}}$,  ${\rm BR} (B_s \to  \mu^+ \mu^-)$, and  various observables of $B \to K^* \mu^+ \mu^-$ and $B_s \to \phi \mu^+ \mu^-$ processes for $b \to s \ell \ell$ transitions for the representative mass of LQ as  $m_{\rm LQ}=1.2$ TeV. Using these constrained couplings, we have predicted the branching ratios of various LFV decays of $B$ and $B_s$ mesons such as $B \to K^{(*)} \ell_i^+ \ell_j^-$,  $B_s \to \phi \ell_i^+ \ell_j^-$ and $B_s \to \ell_i^+ \ell_j^-$. We found that the branching fractions of these decay modes are substantially enhanced due to the effect of $R_2$ SLQ and are within the reach of the Belle-II and LHCb experiments. The observation of these decay modes provide an indirect hint for the existence of the SLQ $R_2(3,2,7/6)$. In addition, we have also investigated the LFV decays  $\Upsilon \to \mu^\pm \tau^\mp$, $\tau \to \mu^- \phi$ and $\tau \to \mu^- \eta (\eta')$. Furthermore, the neutral component of fermion triplet $\Sigma$ contributes to the relic abundance of the Universe near $2.34$ to $2.4$ TeV mass regime. One loop spin-independent DM-nucleon cross section is also suitably obtained within upper limits of experiments such as XNENON1T, LUX and PandaX-II.

\appendix
\acknowledgments 
 SS and RM would like to acknowledge University of Hyderabad IoE project grant no. RC1-20-012. RM acknowledges the support from SERB,
Government of India, through grant No. EMR/2017/001448.
Purushottam Sahu would like to acknowledge the Ministry of Education, Govt of India for financial support. PS also acknowledges the support from the Abdus Salam
International Centre for Theoretical Physics (ICTP) under the 'ICTP Sandwich
Training
Educational Programme (STEP)' SMR.3676 .

\appendix

\section{One loop GUT Threshold corrections to SM gauge couplings}
The analytical relation for the threshold corrections at $M_{GUT}$ in the $G_{SM}$ model,are 
\begin{align}
\pmb{\lambda^{U}_{3C}} = 5 &- 21 \bigg[ 
\eta_{V_4}+ \eta_{V_5} +\eta_{V_6}+\eta_{V_7}+\frac{1}{2} \eta_{V_8}
+\frac{1}{2} \eta_{V_9}
\bigg]
\nonumber \\
&+
2\bigg[
\frac{1}{2}\eta_{S_2} +\frac{1}{2} \eta_{S_3} + \eta_{S_4} + \frac{1}{2} \eta_{S_5}+\frac{15}{2} \eta_{S_7} +\frac{1}{2} \eta_{S_{11}} + \frac{1}{2} \eta_{S_{12}} + \frac{5}{2} \eta_{S_{13}} + \frac{5}{2} \eta_{S_{14}} \nonumber \\
&+\frac{5}{2} \eta_{S_{15}} + \eta_{S_{18}} + \eta_{S_{19}} + \eta_{S_{20}} + 6 \eta_{S_{21}} + 6 \eta_{S_{22}} + \frac{3}{2} \eta_{S_{23}}
\bigg]
 \nonumber \\
&+8\big[\eta_{F_4} + \eta_{F_5} + \eta_{F_6} + \eta_{F_7} + \frac{1}{2}\eta_{F_8} + \frac{1}{2}\eta_{F_9} + 3\eta_{F_{10}}  \big] \\
\pmb{\lambda^{U}_{2L}} = 6 &-21 
\bigg[ \frac{3}{2}\eta_{V_4} +\frac{3}{2} \eta_{V_5}+ \frac{3}{2}\eta_{V_6}+\frac{3}{2} \eta_{V_7}
\bigg] \nonumber \\
&+2 \bigg[\frac{1}{2}\eta_{S_1} +2\eta_{S_6} +12\eta_{S_7} + \frac{1}{2}\eta_{S_{16}} + \frac{1}{2}\eta_{S_{17}} +\frac{3}{2}\eta_{S_{18}} +\frac{3}{2}\eta_{S_{19}} + \frac{3}{2}\eta_{S_{20}} + 4\eta_{S_{21}}+4\eta_{S_{22}}+ 6\eta_{S_{23}}
\bigg]
\nonumber \\
&+8\big[\frac{3}{2}\eta_{F_4} + \frac{3}{2}\eta_{F_5} +\frac{3}{2} \eta_{F_6} + \frac{3}{2}\eta_{F_7} \big] \\
\pmb{\lambda^{U}_{Y}} = 8 &-21
\bigg[\frac{3}{5}\eta_{V_1}+\frac{3}{5} \eta_{V_3} 
+\frac{1}{10}\eta_{V_4}+\frac{5}{2}\eta_{V_5}+\frac{5}{2} \eta_{V_6}+\frac{1}{10}\eta_{V_7}+\frac{4}{5}\eta_{V_8} + \frac{4}{5}\eta_{V_9}
\bigg] \nonumber \\
&+ 2\bigg[
\frac{3}{10}\eta_{S_1}+\frac{1}{5}\eta_{S_2}+\frac{1}{5}\eta_{S_3}+\frac{2}{5}\eta_{S_4}+\frac{1}{5}\eta_{S_5}+\frac{9}{5}\eta_{S_6}+\frac{6}{5}\eta_{S_7}+\frac{3}{5}\eta_{S_9}+\frac{12}{5}\eta_{S_{10}} \nonumber \\
& +\frac{4}{5}\eta_{S_{11}}+\frac{16}{5}\eta_{S_{12}}+\frac{32}{5}\eta_{S_{13}}+\frac{2}{5}\eta_{S_{14}}+\frac{8}{5}\eta_{S_{15}}+\frac{3}{10}\eta_{S_{16}}+\frac{3}{10}\eta_{S_{17}}+\frac{1}{10}\eta_{S_{18}}
 \nonumber\\
 & +\frac{49}{10}\eta_{S_{19}}+\frac{1}{10}\eta_{S_{20}}+\frac{12}{5}\eta_{S_{21}}+\frac{12}{5}\eta_{S_{22}} + \frac{3}{5}\eta_{S_{23}}
\bigg] \nonumber \\
&+8 \bigg[\frac{3}{5}\eta_{F_1} + \frac{3}{5}\eta_{F_3}+ \frac{1}{10}\eta_{F_4}+ \frac{5}{2}\eta_{F_5}+ \frac{5}{2}\eta_{F_6}+ \frac{1}{10}\eta_{F_7}+ \frac{4}{5}\eta_{F_8}+ \frac{4}{5}\eta_{F_9}\bigg]
\end{align}

\clearpage
\bibliography{so10}
\end{document}